\documentclass{aa}  

\usepackage{graphicx}
\usepackage{float}
\usepackage{indentfirst}
\usepackage{lscape}
\usepackage{longtable}
\usepackage{amsmath}
\usepackage{amssymb}
\usepackage{graphicx}
\usepackage{lastpage}
\usepackage{natbib}
\usepackage{txfonts}
\usepackage{xcolor} 
\usepackage[rightcaption]{sidecap}
\usepackage{graphicx} 
\graphicspath{ {images/} }
\usepackage{placeins}
\usepackage{natbib}

\begin{document}
\title{Discovery of linear propadienone:}
\subtitle{Study of the chemistry of linear and cyclic H$_2$C$_3$O and H$_2$C$_3$S in TMC-1}

   \author{G. Esplugues
          \inst{1},          
          J. C. Loison
          \inst{2},          
          M. Agúndez
          \inst{3},
          G. Molpeceres
          \inst{3}, 
          N. Marcelino
          \inst{1, 4},
          B. Tercero  
          \inst{1, 4},
          J. Cernicharo
          \inst{3}
         }

   \institute{Observatorio Astron\'omico Nacional (OAN), Alfonso XII, 3, 28014. Madrid, Spain\\
              \email{g.esplugues@oan.es, jose.cernicharo@csic.es}
       \and
           Institut des Sciences Moléculaires ISM, CNRS UMR 5255, Univ. Bordeaux, 351 Cours de la Libération, F-33400, Talence, France  
        \and
       Instituto de Física Fundamental, CSIC, Calle Serrano 123, E-28006 Madrid, Spain
        \and
Observatorio de Yebes, IGN, Cerro de la Palera s/n, E-19141 Yebes, Guadalajara, Spain  
}

  \abstract   
   {We report the first detection in space of propadienone, the linear isomer (l-H$_2$C$_3$O) of cyclopropenone (c-H$_2$C$_3$O). We also report the first detection of the isotopologues c-H$_2$$^{13}$CCCO and c-HDCCCO of c-H$_2$C$_3$O. The astronomical observations are part of QUIJOTE, a line survey of TMC-1 in the frequency range 31.0-50.3 GHz{\bf{,}} complemented with data between 71.6-116.0 GHz, and carried out with Yebes-40m and IRAM-30m telescopes, respectively. We obtain a total column density of 3.7$\times$10$^{10}$ cm$^{-2}$ for l-H$_2$C$_3$O at an excitation temperature of 4.8 K. We find that the isomer is about eight times less abundant than the cyclic one. We also report a detailed line-by-line study of cyclopropenethione (c-H$_2$C$_3$S) to compare the abundance of the O and S isomers. We find that cyclic O-isomers are more abundant than cyclic S-isomers; however, the opposite trend is found for the most stable linear isomers, with l-H$_2$C$_3$S being more than one order of magnitude more abundant than l-H$_2$C$_3$O.      
A comprehensive theoretical chemical analysis shows that the abundances of the H$_2$C$_3$O and H$_2$C$_3$S isomers are controlled by different formation pathways. 
In particular, while l-H$_2$C$_3$O is potentially produced by dissociative electron recombination reactions, ion-neutral chemistry is more effective at producing l-H$_2$C$_3$S and c-H$_2$C$_3$S.  
}

   \titlerunning{First interstellar detection of propadienone}
   \authorrunning{G. Esplugues et al.}
   \maketitle

\section{\textbf{Introduction}}

Observations of chemical species allow us to constrain the physical conditions of the interstellar medium (ISM), which exhibits a wide range of temperatures from $\sim$10–10$^4$ K, densities from $\sim$10–10$^8$ cm$^{-3}$, and the presence of energetic radiation around newly formed stars, as well as galactic UV radiation at the external layers of the clouds. Among the different molecular species, isomers of a given molecule are particularly important since they have the same constituent atoms and therefore a similar level of chemical complexity. This makes them a powerful tool for probing hypotheses about chemical reaction mechanisms in the ISM as their formation and destruction paths directly affect their abundance ratios \citep[]{Loomis2015}. Many isomers have been detected in the ISM, including some metastable isomers that can be even more abundant than the stable ones \citep[e.g.][]{Halfen2009, Adande2010, Marcelino2010, Cernicharo2022, Cernicharo2024b, Esplugues2024}. This highlights the presence of non-equilibrium chemistry in the ISM.  

Likewise useful are comparisons between chemically analogous isomers, since they might yield important information about their formation pathways. This could be particularly interesting for sulphur (S) as our present understanding of the chemistry of S-bearing species is highly limited by the molecules detected so far and the chemical network used for these species \citep[e.g.][]{Vidal2017, Cernicharo2021a, Cernicharo2021b, Cernicharo2021d, Cernicharo2024b, Cernicharo2024, Esplugues2022, Esplugues2023, Esplugues2025, Cabezas2025, Agundez2025, Fuente2025}. New detected molecules with sulphur could certainly provide new insights, as could comparisons between sulphur-bearing molecules and their oxygen counterparts, since both elements share the same electron configurations \citep[]{Remijan2025, Cabezas2025, Agundez2025}.

To this end, we analyse and compare the emission of H$_2$C$_3$O and H$_2$C$_3$S in TMC-1. In particular, propadienthione (l-H$_2$C$_3$S) and cyclopropenethione (c-H$_2$C$_3$S) have recently been detected in TMC-1 by \cite{Cernicharo2021a} and \cite{Remijan2025}, respectively. Cyclopropenone (c-H$_2$C$_3$O) was first observed in Sgr\,B2 by \cite{Hollis2006}, and its isomer propynal (HCCCHO) was first detected in TMC-1 by \cite{Irvine1988}. However, l-H$_2$C$_3$O remained undetected in spite of being the most thermodynamically stable of the three isomers \citep[]{Komornicki1981, Loomis2015, Shingledecker2019}.  

In this paper, we provide the first detection of propadienone (l-H$_2$C$_3$O) in the ISM, as well as the deuterated (c-HDC$_3$O) and $^{13}$C substitutions of cyclopropenone. Observations are described in Sect. \ref{Observations}. We also present and discuss our results in Sects. \ref{section:results} and \ref{section:discussion}, respectively, including results from a chemical model that allow us to better compare the chemical reactions forming and destroying these S- and O-isomers. We finally summarise our conclusions in Sect. \ref{section:summary}.

\section{Observations}
\label{Observations}

The observations presented in this work are derived from the ongoing QUIJOTE line survey (31.0-50.3 GHz) carried out with the Yebes 40m telescope, complemented with data (71.6-116.0 GHz) observed with the IRAM 30m telescope. A detailed description of the QUIJOTE line survey and the data-analysis procedure are provided in \cite{Cernicharo2021c}. Observations were carried out at the cyanopolyyne peak of TMC-1 ($\alpha$$_{\mathrm{J2000}}$ = 04$^{\mathrm{h}}$:41$^{\mathrm{m}}$:41.9$^{\mathrm{s}}$,  $\delta$$_{\mathrm{J2000}}$ = 25$^{\mathrm{o}}$:41$^{\mathrm{'}}$:27.0$^{\mathrm{''}}$). 

The Yebes 40m survey was performed in several sessions between 2019 and 2024, using new receivers built within the Nanocosmos project\footnote{https://nanocosmos.iff.csic.es/} \citep[]{Tercero2021}. The Q-band receiver consists of two cold high-electron-mobility transistor (HEMT) amplifiers that cover the 31.0-50.3 GHz (7 mm) Q band with horizontal and vertical polarizations. Fast Fourier transform spectrometers (FFTSs) with 8$\times$2.5 GHz bandwidth and a spectral resolution of 38.15 kHz ($\sim$0.27 km s$^{-1}$) were used. The observational mode was frequency-switching with frequency throws of either 10 or 8 MHz. In addition, different central frequencies were used during the runs to ensure that no spurious spectral ghosts were produced in the down-conversion chain. The total observing time on the source for the data taken with the frequency throws of 10 MHz and 8 MHz is 772.6 and 736.6 hours, respectively. Hence, the total observing time on the source is 1509.2 hours. The QUIJOTE sensitivity varies between 0.06 mK at 32 GHz and 0.18 mK at 49.5 GHz \citep[]{Cernicharo2024c}.  

The observations from the IRAM 30m telescope, located in Pico Veleta (Granada, Spain), were carried out using frequency-switching observing mode with a frequency throw of 7.14 MHz \citep[]{Cernicharo2012b, Marcelino2023, Cernicharo2024b}, the broadband EMIR (E090) receiver covering the frequency range 71.6-117.6 GHz, and the FFTSs with 48 kHz spectral resolution. 
Observations were performed in several runs between 2012 and 2021. Each frequency setup was observed for $\sim$2 hr, with pointing checks in between on strong and nearby quasars. Pointing errors were always within 3$\arcsec$.

The intensity scale in antenna temperature ($T$$^{\star}_{\mathrm{A}}$), corrected for atmospheric absorption and for antenna ohmic and spillover losses, was calibrated using two absorbers at different temperatures as well as the atmospheric transmission model \citep[ATM;][]{Cernicharo1985b, Pardo2025}. The absolute calibration uncertainty is 9$\%$.
The data were reduced and processed by using the CLASS package provided within the GILDAS software\footnote{http://www.iram.fr/IRAMFR/GILDAS}, developed by the IRAM Institute.

\section{Results}
\label{section:results}

Regarding linear (l-) isomers, we detected for the first time in the ISM two rotational transitions in band Q of l-H$_2$C$_3$O (4$_{0}$$_{,}$$_{4}$-3$_{0}$$_{,}$$_{3}$ and 5$_{0}$$_{,}$$_{5}$-4$_{0}$$_{,}$$_{4}$), with $E$$_{\mathrm{u}}$=4.3 K and 6.4 K, respectively. These transitions have the same frequency for both the unresolved ortho(o-) and para(p-). 
We used the Cologne Database for Molecular Spectroscopy \citep[CDMS,][]{Muller2001, Muller2005, Endres2016} and the MADEX catalogue \citep[]{Cernicharo2012}, which considers spectroscopic laboratory data for propadienone from \cite{Brown1981}. The lines exhibit antenna temperatures of 0.44 and 0.56 mK, as well as line-widths of 0.54 to 1.1 km s$^{-1}$, similar to those found for other sulphur species in TMC-1 \citep[e.g.][]{Cernicharo2024, Agundez2025, Esplugues2025}. Figure \ref{figure:o-H2CCCO_lines}\footnote{Quantum nunmbers are indicated in each panel included in Figures \ref{figure:o-H2CCCO_lines}-\ref{figure:HC13CCHO_lines} of the Appendix. The red line indicates the LTE synthetic spectrum from a fit to the observed line profiles. The horizontal green line indicates the 1$\sigma$ noise level.} shows the observed line profiles (black lines) of l-H$_2$C$_3$O. See also Table \ref{table:line_parameters}, which lists the line parameters obtained from Gaussian fits of the detected lines. We also detected eleven lines of HCCCHO (Fig. \ref{figure:HCCCHO_lines}); one line of its isotopologue, H$^{13}$CCCHO (Fig. \ref{figure:H13CCCHO_lines}); and another line of HC$^{13}$CCHO (Fig. \ref{figure:HC13CCHO_lines}). For the S counterpart, we detected seven lines of o-H$_2$C$_3$S (Fig. \ref{figure:o-H2CCCS_lines}) and three lines of p-H$_2$C$_3$S (Fig. \ref{figure:p-H2CCCS_lines}), with $E$$_{\mathrm{u}}$ ranging from 6.5 K to 10.9 K, and linewidths ranging from 0.58-0.77 km s$^{-1}$ (Table \ref{table:line_parameters}). These detections rely on longer observational time than the original detection by \cite{Cernicharo2021a}; therefore, the data shown in this paper (Tables \ref{table:N_sources}, \ref{table:OPRs}, and \ref{table:line_parameters}) should be considered as the reference data for this molecule.

Regarding cyclic (c-) isomers, we detected (Table \ref{table:line_parameters}) six lines of o-c-H$_2$C$_3$O (Fig. \ref{figure:o-c-H2C3O_lines}), three lines of p-c-H$_2$C$_3$O (Fig. \ref{figure:p-c-H2C3O_lines}), seven lines of o-c-H$_2$C$_3$S (Fig. \ref{figure:o-c-H2CCCS_lines}), and three lines of p-c-H$_2$C$_3$S (Fig. \ref{figure:p-c-H2CCCS_lines}), with $E$$_{\mathrm{u}}$ ranging from 3.2 K to 17.2 K. They all were observed at 7 mm, except for four lines of o-c-H$_2$C$_3$O that were detected at 3 mm. The lines exhibit antenna temperatures between 0.32 and 23.6 mK (with lines from c-H$_2$C$_3$O being significantly more intense than those from c-H$_2$C$_3$S). Linewidths range from 0.49 to 0.91 km s$^{-1}$. The c-H$_2$C$_3$O lines are not blended with other species; however, four lines of c-H$_2$C$_3$S are contaminated by emission from other molecules (see Table \ref{table:line_parameters}). This blending of lines for intensities below 1 mK should be be considered as a warning for blind statistical stacking detection methods. We also detected for the first time in the ISM the deuterated molecule c-HDC$_3$O by observing two rotational lines (see Table \ref{table:line_parameters} and Fig. \ref{figure:c-HDC3O_lines}, indicating c-HDC$_3$O lines with $E$$_{\mathrm{u}}$$<$15 K and $A$$_{\mathrm{ul}}$$>$10$^{-6}$ s$^{-1}$). For c-H$_2$C$_3$O, we also detected some isotopologues. In particular, we observed two lines of c-H$_2$$^{13}$CCCO (3$_{0}$$_{,}$$_{3}$-2$_{0}$$_{,}$$_{2}$ and 3$_{1}$$_{,}$$_{2}$-2$_{1}$$_{,}$$_{1}$; see Fig. \ref{figure:c-H213CCCO_lines}) above 3.5$\sigma$ without contamination from any other spectral feature. We also detected for the first time one line from c-H$_2$CC$^{13}$CO (6$_{1}$$_{,}$$_{5}$-5$_{1}$$_{,}$$_{4}$; see Fig. \ref{figure:o-c-H2CC13CO_lines}). However, since it is the only one detected, and its line intensity is $\sim$3$\sigma$, we considered it a tentative detection.

\subsection{Rotational diagrams}
\label{section:rotational_diagrams}

For the molecules with more than two detected lines, we computed a representative rotational temperature ($T$$_{\mathrm{rot}}$) by constructing a rotational diagram, assuming a single rotational temperature for all the energy levels \citep[]{Goldsmith1999}. 
The resulting rotational diagrams are shown in Fig. \ref{figure:RD}, and the obtained rotational temperatures (with the uncertainties obtained from the least-squares fit) are listed in Table \ref{table:N_sources}. 
We did not apply the rotational diagram method to c-HDC$_3$O, since we had only two  rotational transitions. Therefore, for c-HDC$_3$O, we assumed the $T$$_{\mathrm{rot}}$ obtained for o-c-H$_2$C$_3$O. For the isotopologue c-H$_2$$^{13}$CCCO (with two detected lines), we also adopted the $T$$_{\mathrm{rot}}$ from o-c-H$_2$C$_3$O. Since we only observed one line for H$^{13}$CCCHO and HC$^{13}$CCHO, we adopted $T$$_{\mathrm{rot}}$ from HCCCHO. Since only one line (6$_{0}$$_{,}$$_{6}$-5$_{0}$$_{,}$$_{5}$) of p-c-H$_2$C$_3$S is $>$4$\sigma$ and is not blended, we assumed the same $T$$_{\mathrm{rot}}$ than that derived for o-c-H$_2$C$_3$S.

Most rotational temperatures range from 4.8-6.5 K. Only o-c-H$_2$C$_3$S and HCCCHO have slightly higher values ($\sim$8 K), but large uncertainties (Fig. \ref{figure:RD}). We also note that lines of o-c-H$_2$C$_3$S (5$_{3}$$_{,}$$_{3}$-4$_{3}$$_{,}$$_{2}$ and 5$_{3}$$_{,}$$_{2}$-4$_{3}$$_{,}$$_{1}$) are blended with the hyperfine structure (hfs) of H$^{13}$CCNC, which increases the uncertainty on the fitted $T$$_{\mathrm{rot}}$.

\subsection{Column densities and abundances}
\label{section:column_densities_abundances}

In order to calculate the column densities of the observed H$_2$C$_3$O and H$_2$C$_3$S isomers, we used the MADEX code assuming LTE (Local Thermodynamic Equilibrium) at the rotational temperatures derived above due to the lack of collisional rates. 
In our case, we used the $T$$_{\mathrm{rot}}$ from the transition analysis listed in Table \ref{table:N_sources}. The use of this code allows us to compare the synthetic line profiles with observed ones. 

The MADEX code provides antenna temperature intensities for each line after taking the different telescope beams and efficiencies into account. We assumed a diameter size of 80$\arcsec$ \citep[]{Fosse2001}, compatible with the emission size of most of the molecules mapped in TMC-1 \citep[]{Cernicharo2023}. For each molecule, we considered the average value of linewidths obtained from the Gaussian fits (Table \ref{table:line_parameters}). We left the column density as the only free parameter. Figures \ref{figure:o-H2CCCO_lines}-\ref{figure:HC13CCHO_lines} show the synthetic spectra from the LTE models in red, overlaid with the observed line profiles (black). The derived column densities (with estimated errors of $\sim$20$\%$) are listed in Table \ref{table:N_sources}. Abundances for each molecule, also shown in Table \ref{table:N_sources}, were derived considering an H$_2$ column density of 10$^{22}$ cm$^{-2}$ for TMC-1 \citep[]{Cernicharo1987}. For those species that were undetected (e.g. H$^{13}$CCCHO), or for which only detected one line (e.g. o-c-H$_{2}$CC$^{13}$CO), we provided upper limits. To derive these upper limits for each species, once the physical parameters were fixed, we varied the column density until the model fit reached the intensity peak of any observed line. We did not allow the model fit to be greater than any observed line.

\section{Discussion}
\label{section:discussion}

We detect for the first time in space propadienone (l-H$_2$C$_3$O). This is the most stable isomer of H$_2$C$_3$O, with the lowest binding energy among them \citep[]{Loomis2015}. However, despite this, it remained undetected until now. In particular, we have derived its abundance to be 3.7$\times$10$^{-12}$. The abundances derived for c-H$_2$C$_3$O and HCCCHO are 3.2$\times$10$^{-11}$ and 9.2$\times$10$^{-11}$, respectively. 
This implies that propadienone is one order of magnitude less abundant than c-H$_2$C$_3$O and HCCCHO. Nevertheless, this is not the first case where the least stable isomers were detected first and exhibited higher abundances than the most stable ones \citep[see e.g.][]{Brunken2009, Marcelino2010, Halfen2009, Esplugues2024}, suggesting that kinetic effects have a more important role than thermodynamic effects on the chemical formation pathways of these isomers. It therefore highlights the importance of knowing the formation path of the different isomers independently of their relative energies. The principle of minimum energy \citep[]{Lattelais2009} does not hold under all conditions of interstellar clouds.    

\begin{table}
\centering
\caption{Obtained column densities, $N$ (cm$^{-2}$), and abundances ($X$).}
\begin{center}
\begin{tabular}{llllll}
\hline 
\hline
Molecule               & $T$$_{\mathrm{rot}}$ (K)   & $N$  (cm$^{-2}$)                  &  $X$$^*$     \\
\hline
\hline 
l-H$_2$C$_3$O          &  4.8$^a$               & (3.7$\pm$0.7)$\times$10$^{10}$    &  (3.7$\pm$0.7)$\times$10$^{-12}$   \\
\hline
o-c-H$_2$C$_3$O        &  4.8$\pm$1.1             & (2.3$\pm$0.5)$\times$10$^{11}$    &  (2.3$\pm$0.5)$\times$10$^{-11}$   \\
p-c-H$_2$C$_3$O        &  6.5$\pm$0.6             & (8.6$\pm$1.7)$\times$10$^{10}$    &  (8.6$\pm$1.7)$\times$10$^{-12}$   \\ 
\hline
c-HDC$_3$O             &  4.8$^a$          & (2.6$\pm$0.5)$\times$10$^{10}$    &  (2.6$\pm$0.5)$\times$10$^{-12}$   \\
\hline
c-H$_2$$^{13}$CCCO     &  4.8$^a$           & (1.4$\pm$0.3)$\times$10$^{10}$    &  (1.4$\pm$0.3)$\times$10$^{-12}$  \\
c-H$_2$CC$^{13}$CO     &  4.8$^a$           & $<$(5.6$\pm$1.1)$\times$10$^{10}$ &  $<$(5.6$\pm$1.1)$\times$10$^{-12}$  \\
\hline
o-H$_2$C$_3$S          &  5.6$\pm$1.5               & (6.0$\pm$1.2)$\times$10$^{11}$    &  (6.0$\pm$1.2)$\times$10$^{-11}$  \\
p-H$_2$C$_3$S          &  5.5$\pm$0.6               & (1.3$\pm$0.3)$\times$10$^{11}$    &  (1.3$\pm$0.3)$\times$10$^{-11}$   \\
\hline
o-c-H$_2$C$_3$S        &  8.2$\pm$3.5               & (3.8$\pm$0.8)$\times$10$^{10}$    &  (3.8$\pm$0.8)$\times$10$^{-12}$     \\
p-c-H$_2$C$_3$S        &  8.2$^b$                   & (9.5$\pm$1.9)$\times$10$^{9}$     &  (9.5$\pm$1.9)$\times$10$^{-13}$    \\
\hline
HCCCHO                 &  8.1$\pm$2.1               & (9.2$\pm$2.0)$\times$10$^{11}$    &  (9.2$\pm$2.0)$\times$10$^{-11}$    \\
\hline
H$^{13}$CCCHO          &  8.1$^c$               & $<$(4.4$\pm$0.8)$\times$10$^{10}$ &  $<$(4.4$\pm$0.8)$\times$10$^{-12}$    \\
HC$^{13}$CCHO          &  8.1$^c$               & $<$(3.8$\pm$0.7)$\times$10$^{10}$ &  $<$(3.8$\pm$0.7)$\times$10$^{-12}$    \\
HCC$^{13}$CHO          &  8.1$^c$               & $<$(2.3$\pm$0.5)$\times$10$^{10}$ &  $<$(2.3$\pm$0.5)$\times$10$^{-12}$    \\
\hline
HCCCHS$^d$             &  5.0$\pm$0.5               & (3.2$\pm$0.4)$\times$10$^{11}$ &  (3.2$\pm$0.4)$\times$10$^{-11}$    \\
\hline
\end{tabular}
\label{table:N_sources}
\end{center}
\tablefoot{
\tablefoottext{*}{Assuming an H$_2$ column density of 10$^{22}$ cm$^{-2}$ for TMC-1 (Cernicharo $\&$ Guelin 1987).}
\tablefoottext{a}{Adopted from o-c-H$_2$C$_3$O.}
\tablefoottext{b}{Adopted from o-c-H$_2$C$_3$S.}
\tablefoottext{c}{Adopted from HCCCHO.}
\tablefoottext{d}{Adopted from \cite{Cernicharo2021b}.}
}
\end{table}

Particularly interesting is the comparison between H$_2$C$_3$S and its oxygen analogue H$_2$C$_3$O, given that oxygen and sulphur share similar electron configurations.
While we derived an abundance of $X$=3.7$\times$10$^{-12}$ for l-H$_2$C$_3$O, we obtained $X$(l-H$_2$C$_3$S)=7.3$\times$10$^{-11}$ (Table \ref{table:N_sources}) for its sulphur counterpart, i.e. l-H$_2$C$_3$S is about twenty times more abundant than l-H$_2$C$_3$O. By contrast, we obtained the opposite trend for the cyclic isomers, with abundances of 3.2$\times$10$^{-11}$ and 4.8$\times$10$^{-12}$ for c-H$_2$C$_3$O and c-H$_2$C$_3$S, respectively. This means that the oxygen cyclic isomer is about seven times more abundant than the sulphur cyclic one. For the other linear isomers, HCCCHO and HCCCHS, we also find that the oxygen one is about three times more abundant than the sulphur one. 
These results suggest that, for the metastable isomers, those containing oxygen are more abundant than their sulphur analoguess. However, for the stable isomers, the opposite trend is found.

\subsection{Isotopic fractionation}
\label{section:isotopic_fractionation}

We also detected for the first time in space the isotopologue c-H$_2$$^{13}$CCCO, with the clear observation of two transitions (above 3.5$\sigma$). We derived $N$(c-H$_2$$^{13}$CCCO)=(1.4$\pm$0.3)$\times$10$^{10}$ cm$^{-2}$, which implies $^{12}$C/$^{13}$C=23$\pm$9. For c-H$_2$CC$^{13}$CO, we only detected one transition; therefore, as previously stated, we considered it a tentative detection, with $N$(c-H$_2$CC$^{13}$CO)$<$(5.6$\pm$1.1)$\times$10$^{10}$ cm$^{-2}$. For the case of propynal, we did not detected any $^{13}$C isotopologue; however, we provided upper limits (Table \ref{table:N_sources}) instead. From these results, we derived a HCCCHO/H$^{13}$CCCHO ratio $>$20.9 in agreement with the result obtained for the c-H$_2$C$_3$O/c-H$_2$$^{13}$CCCO ratio. These $^{12}$C/$^{13}$C values are significantly different from that of the Solar System value \citep[$^{12}$C/$^{13}$C=89,][]{Wilson1994} and from the standard value of $\sim$70 in the local ISM \citep[e.g.][]{Wilson1994, Ritchey2011}. Comparing with recent results found for TMC-1, \cite{Tercero2024} obtained $^{12}$C/$^{13}$C$\sim$69-106 using HCCCN and deduced a clear fractionation for the species with the $^{13}$C isotope adjacent to nitrogen. \cite{Cernicharo2024d} obtained $^{12}$C/$^{13}$C$\sim$88-94 and $\sim$93-100 though the isomers HNCCC and HCCNC, respectively, but without finding any isotopic fractionation trend depending on the $^{13}$C position. This suggests chemical differences in the processes governing isotopic fractionation for isomerism.
Also in TMC-1, \cite{Navarro-Almaida2023} obtained significantly low $^{12}$C/$^{13}$C values (16-23) using HCN and H$^{13}$CN. These values are similar to those deduced by \cite{Daniel2013} ($\sim$20-30) and \cite{Magalhaes2018} ($\sim$45) using the same molecules, as well as HNC and CN, towards the pre-stellar core B1\,b and the starless core L\,1498, respectively. Nevertheless, these molecules are likely affected by opacity effects, which strongly impact the derivations of column densities and may lead to an underestimation of the main isotope column density, and thus of the $^{12}$C/$^{13}$C ratios.   
Lines of o-c-H$_2$C$_3$O are, however, optically thin, suggesting that it is enriched in $^{13}$C in TMC-1. C fractionation and, in particular, $^{13}$C enhancement in dense interstellar clouds have been studied through observations and distinct time-dependent chemical models \citep[e.g.][]{Langer1984, Langer1990, Furuya2011, Roueff2015}. In particular, \cite{Langer1984} introduced different isotopic exchange reactions in their model, suggesting that C fractionation in the ISM results from the chemical reaction,

\begin{equation}
\mathrm{^{13}C^+ + {^{12}CO} \leftrightarrow {^{13}CO} + {^{12}C^+} + 35 K}, 
\label{equation:1}
\end{equation}

\noindent and concluded that the lower the temperature, the higher the chemical fractionation of C-bearing species. It was also deduced that when CO molecules become the main reservoir of carbon, although the $^{13}$C concentration is low, reaction (1) still leads to a small $^{13}$CO enrichment \citep[]{Langer1990, Milam2005}, while other carbon-containing species (for instance, CH and other carbon chains) become depleted in $^{13}$C. This was also found by \cite{Roueff2015} when studying C fractionation for C, CH, CO, and HCO$^+$. They found that HCO$^+$ is marginally enriched in $^{13}$C at steady state, whereas HCN and HNC are significantly depleted in $^{13}$C. These results show how the $^{12}$C/$^{13}$C ratio can vary among different chemical species.

In this paper, we also report the detection of the deuterated version of cyclopropenethione, i.e. c-HDC$_3$O, with $N$=2.6$\times$10$^{10}$ cm$^{-2}$. It implies an H/D ratio of 12$\pm$5. This result is similar to the ratios found in TMC-1 for CH$_3$CN/CH$_2$DCN=11, CH$_3$CCH/CH$_2$DCCH=10, and H$_2$CCN/HDCCN=20 \citep[]{Cabezas2021}. However, it also significantly differs from other H/D ratios, such as C$_4$H/C$_4$D=118 \citep[]{Cabezas2021} and HC$_5$N/DC$_5$N=82 \citep[]{Cernicharo2020} obtained from TMC-1. Nevertheless, theoretical chemical models predict that the H/D values for these two ratios are $\sim$20-55 \citep[]{Cabezas2022}, which agree better with our results. In any case, a more comprehensive deuteration study of large carbon-chain molecules should be carried out in order to understand the large H/D observational differences between similar species of the same region.

\subsection{The ortho-to-para ratio}
\label{section:ortho-para_ratio}

Molecular hydrogen (H$_2$) has two nuclear spin isomers, ortho and para, in which the two nuclear spins of protons are parallel and antiparallel, respectively. The ortho-to-para ratio (OPR) is known to affect chemical evolution as well as gas dynamics in space \citep[e.g.][]{Faure2019, Tsuge2021, Yocum2023}. In thermodynamic equilibrium, OPR (with a statistical weight value of 3) is determined by the reaction heat released compared to the energy difference between the lowest ortho and para states \citep[]{Minh1991}. If molecules are produced with excess energy by ion-molecule reactions, the ratio is expected to be that of the statistical weight \citep[]{Kawaguchi1991}. Nevertheless, factors related to formation processes, such as the spin state of precursor molecules and the structure of the reaction energy surface, might constrain the OPR to a non-equilibrium value. 

In this paper, we report several ortho- and para- transitions of the cyclic isomers c-H$_2$C$_3$O and c-H$_2$C$_3$S, and of the linear isomer l-H$_2$C$_3$S (Table \ref{table:line_parameters}). In particular, we obtained OPRs of 2.7$\pm$1.0 and 4.0$\pm$1.6 for c-H$_2$C$_3$O and c-H$_2$C$_3$S, respectively, and OPR(H$_2$C$_3$S) = 4.6$\pm$1.9, in agreement with that found by \cite{Cernicharo2021a} (see Table \ref{table:OPRs}). These OPRs are compatible with the 3/1 value expected from the statistical spin degeneracies. Comparing with other OPR from carbon chain molecules in TMC-1, the obtained OPR values for H$_2$CCS and c-C$_3$H$_2$ are 3.3$\pm$0.7 and 3.2$\pm$0.3, respectively \citep[]{Cernicharo2021e, Cernicharo2021a}. Similar values were also found for other sulphur carbon molecules in a sample of starless cores at Taurus and Perseus, such as OPR(H$_2$CS) = 2.4$\pm$0.9 \citep[]{Esplugues2022}.

\begin{table}
\centering
\caption{OPRs for carbon chain molecules.}
\begin{center}
\begin{tabular}{lll}
\hline 
\hline
Molecule           & OPR               &  References              \\
\hline
\hline 
H$_2$CS             &  2.4$\pm$0.9      & \citep[]{Esplugues2022}      \\ 
H$_2$CCS            &  3.3$\pm$0.7      & \citep[]{Cernicharo2021a}      \\ 
H$_2$C$_3$S         &  4.6$\pm$0.8      & \citep[]{Cernicharo2021a}      \\ 
c-C$_3$H$_2$        &  3.2$\pm$0.3      & \citep[]{Cernicharo2021e}      \\ 
c-H$_2$C$_3$O       &  2.7$\pm$1.0      & This paper      \\
c-H$_2$C$_3$S       &  4.0$\pm$1.6      & This paper      \\
H$_2$C$_3$S         &  4.6$\pm$1.9      & This paper      \\
\hline
\end{tabular}
\label{table:OPRs}
\end{center}
\end{table}

\subsection{Chemical modelling}
\label{section:chemical_mechanisms_H2CCCO}

To carry out a theoretical study of the chemistry involved in the formation and destruction of the isomers of H$_2$C$_3$O and H$_2$C$_3$S, we used a model based on the Nautilus code \citep[]{Ruaud2016}, a three-phase (gas, dust grain ice surface, and dust grain ice mantle) time-dependent chemical model employing kida.uva.2024 \citep[]{Wakelam2024} as the basic reaction network. This model was recently updated for a better description of COM (Complex Organic Molecules) chemistry on interstellar dust grains and in the gas-phase \citep[]{Manigand2021, Coutens2022}. There are 800 individual species included in the network that are involved in approximately 9000 separate reactions. Elements are either initially in their atomic or ionic forms in this model (elements with an ionisation potential $<$ 13.6 eV are considered to be fully ionised), and the C/O elemental ratio is equal to 0.71 in this work. The initial simulation parameters are the same as those in Table 4 of \cite{Hickson2024} except for the initial oxygen atom, for which we used a depleted abundance (O/H=1.4$\times$10$^{-4}$), and for Sulphur (S$^{+}$/H=3.0$\times$10$^{-6}$), giving much better results. The grain surface and the mantle are both chemically active for these simulations, while accretion and desorption were only allowed between the surface and the gas-phase. The dust-to-gas ratio (in terms of mass) is 0.01. A sticking probability of 1 was assumed for all neutral species while desorption occurs by both thermal and non-thermal processes (cosmic rays, chemical desorption) including sputtering of ices by cosmic-ray collisions \citep[]{Wakelam2021}.

\begin{figure}
\centering
\includegraphics[scale=0.79, angle=0]{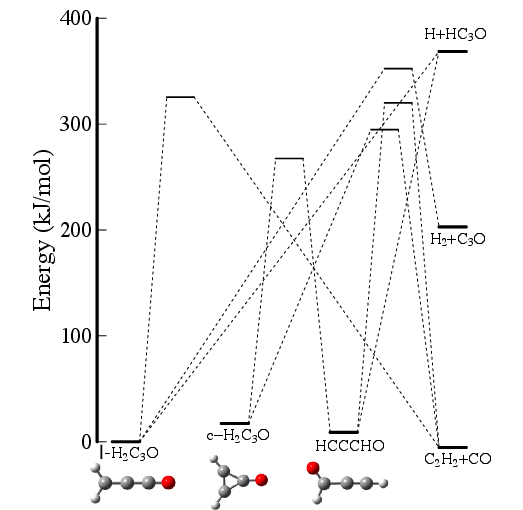}  \hspace{0.3cm}
\caption{Simplified diagram (without HCCHCO) of potential energy for H$_2$C$_3$O isomers calculated at the M06-2X/AVTZ level.}
\label{figure:energy_diagramas_H2C3O}
\end{figure}

The chemistry of H$_2$C$_3$O isomers, including c-H$_2$C$_3$O, was reviewed in \cite{Loison2016b}, but the chemistry for H$_2$C$_3$S was not. Here, we complement that study by reviewing the chemistry of H$_2$C$_3$S isomers, as well as the chemistry of carbon chains in clouds, in relation to the sulphur chemistry studies by \cite{Loison2016b} and \cite{Vidal2017}. In our current chemical network, propynal (HCCCHO) is produced efficiently in the gas phase through the O + C$_3$H$_3$ reaction, while cyclopropenone (c-H$_2$C$_3$O) is produced relatively efficiently through the OH + c-C$_3$H$_2$ reaction. For propadienone (l-H$_2$C$_3$O), a production pathway is the OH + t-C$_3$H$_2$ reaction, as in \cite{Lee2006} and \cite{Loison2016b}. 
Even if this last reaction is limited by the low t-C$_3$H$_2$ abundance, the detection of l-H$_2$C$_3$O is a potential proxy for estimating the abundance of t-C$_3$H$_2$. However, another potential production pathway for l-H$_2$C$_3$O production, as well as the other H$_2$C$_3$O isomers, is the dissociative electron recombination (DR) of HCCCHOH$^+$ and C$_2$H$_3$CO$^+$. These two ions are produced by the protonation of H$_2$C$_3$O isomers, with C$_2$H$_3$CO$^+$ also being produced by the radiative association of C$_2$H$^{+}_{\mathrm{3}}$ + CO \citep[]{Herbst1984, Scott1995, Maclagan1995}. These DR were neglected as H$_2$C$_3$O isomers pathways in \cite{Loison2016b} in favour of the production of C$_2$H$_2$ + CO. Indeed, given the very high exothermicity of DR (431 kJ/mol for the C$_2$H$_3$CO$^+$ + e$^-$ $\rightarrow$  H$_2$C$_3$O + H, and 558 kJ/mol for the HCCCHOH$^+$ + e$^-$ $\rightarrow$  HCCCHO + H), H$_2$C$_3$O isomers will be produced partly above their isomerisation barriers and also above their dissociation barriers as shown in Fig. \ref{figure:energy_diagramas_H2C3O}. 
In this diagram, we do not show the HCCHCO species, which is a fourth isomer that is much less stable, possessing a very shallow potential well that easily yields C$_2$H$_2$ + CO. It is involved in the dissociation of l-H$_2$C$_3$O and c-H$_2$C$_3$O. In the current version of our chemical model, the main route of evolution of the H$_2$C$_3$O isomers produced through the DR of HCCCHOH$^+$ and C$_2$H$_3$CO$^+$ is the dissociation into C$_2$H$_2$ + CO. However, an experimental study of the branching ratio of HCCCHOH$^+$ and C$_2$H$_3$CO$^+$ is necessary. These production routes are necessarily in the minority for HCCCHO and c-H$_2$C$_3$O, but may not be so for l-H$_2$C$_3$O. Indeed, a branching ratio of 5$\%$ is competitive for the production of l-H$_2$C$_3$O compared to the OH + t-C$_3$H$_2$ pathway.

\begin{figure}
\centering
\includegraphics[scale=0.38, angle=0]{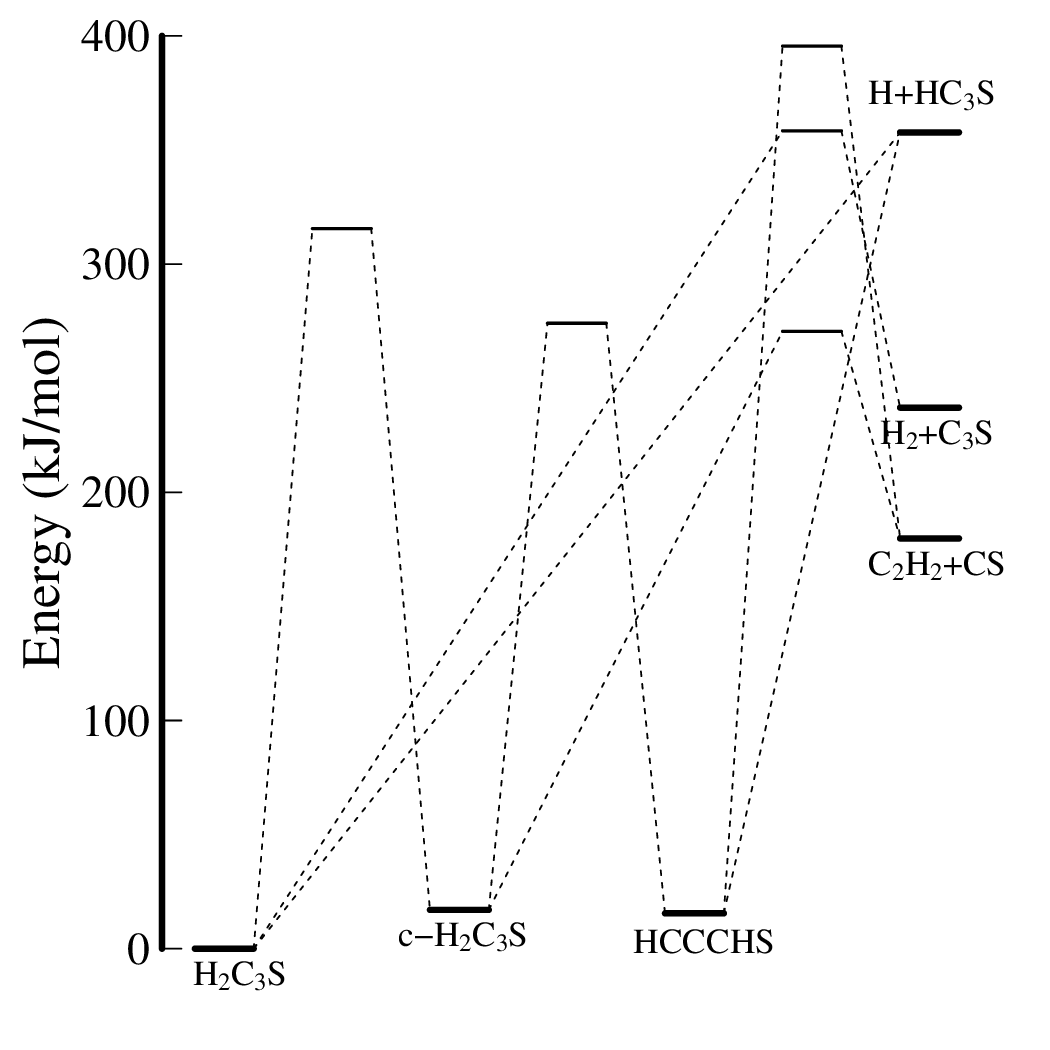}  \hspace{0.3cm}
\caption{Diagram of potential energy for H$_2$C$_3$S isomers calculated at the M06-2X/AVTZ level.}
\label{figure:energy_diagramas_H2C3S}
\end{figure}

Although the chemistry of H$_2$C$_3$S isomers might show similarities to that of H$_2$C$_3$O, there are also many differences among them. HCCCHS is produced mainly and efficiently by the S + C$_3$H$_3$ similar to HCCCHO. However, the production pathways of l-H$_2$C$_3$S and c-H$_2$C$_3$S differ from those of oxygenated analogues because the abundance of SH is much lower than that of OH. Therefore, SH + C$_3$H$_2$ reactions are not very efficient even though the SH + c-C$_3$H$_2$ reaction is a pathway for producing c-H$_2$C$_3$S \citep[]{Remijan2025}. Ionic chemistry is more effective at producing l-H$_2$C$_3$S and c-H$_2$C$_3$S than oxygenated equivalents. This is partly because the abundance of S$^+$ is much higher than the abundance of O$^+$, and the reactions S$^+$ + CH$_3$CCH and S$^+$ + CH$_2$CCH$_2$ efficiently produce C$_2$H$_3$CS$^+$ \citep[]{Smith1988, Decker2001}. Furthermore, the CS bond energy is lower than the CO bond energy. Therefore, the pathway leading to C$_2$H$_2$ + CS is less favoured during the DR of C$_2$H$_3$CS$^+$ than C$_2$H$_2$ + CO during the DR of C$_2$H$_3$CO$^+$ (see Figs. \ref{figure:energy_diagramas_H2C3O} and \ref{figure:energy_diagramas_H2C3S}). Another minor but sulphur-specific pathway in the gas phase for l-H$_2$C$_3$S production is the CH$_3$ + C$_2$S reaction. The oxygenated equivalent, CH$_3$ + C$_2$O, is certainly just as fast at low temperatures, but preferentially produces C$_2$H$_3$ + CO. The differences in the production efficiency of H$_2$C$_3$S compared to H$_2$C$_3$O clearly explain their respective abundances and the fact that H$_2$C$_3$S is significantly more abundant than H$_2$C$_3$O. As with oxygenated equivalents, the branching ratios of the DR of C$_2$H$_3$CS$^+$ and HCCCHSH$^+$ (whose DR rates are shown in Table \ref{table:rates}) are unknown. The lack of knowledge about these branching ratios leads to considerable uncertainty about the modelled abundances of l-H$_2$C$_3$S and c-H$_2$C$_3$S.

\begin{SCfigure*}[0.5][h]
\caption{Abundances of H$_2$C$_3$O and H$_2$C$_3$S isomers as a function of time predicted by our models with $n$(H$_2$)=2$\times$10$^4$ cm$^{-3}$. The horizontal rectangles represent the observed abundances of HCCCHO \citep[]{Cernicharo2020b}, c-H$_2$C$_3$O \citep[]{Cernicharo2020b}, l-H$_2$C$_3$O (this work), HCCCHS \citep[]{Cernicharo2021a}, c-H$_2$C$_3$S \citep[]{Remijan2025}, H$_2$C$_3$S \citep[]{Cernicharo2021a}. The vertical rectangle represents the chemical age given by minimizing the difference between the observations for TMC-1 and the model for 62 species \citep[]{Wakelam2006}.}
\includegraphics[width=0.6\textwidth]{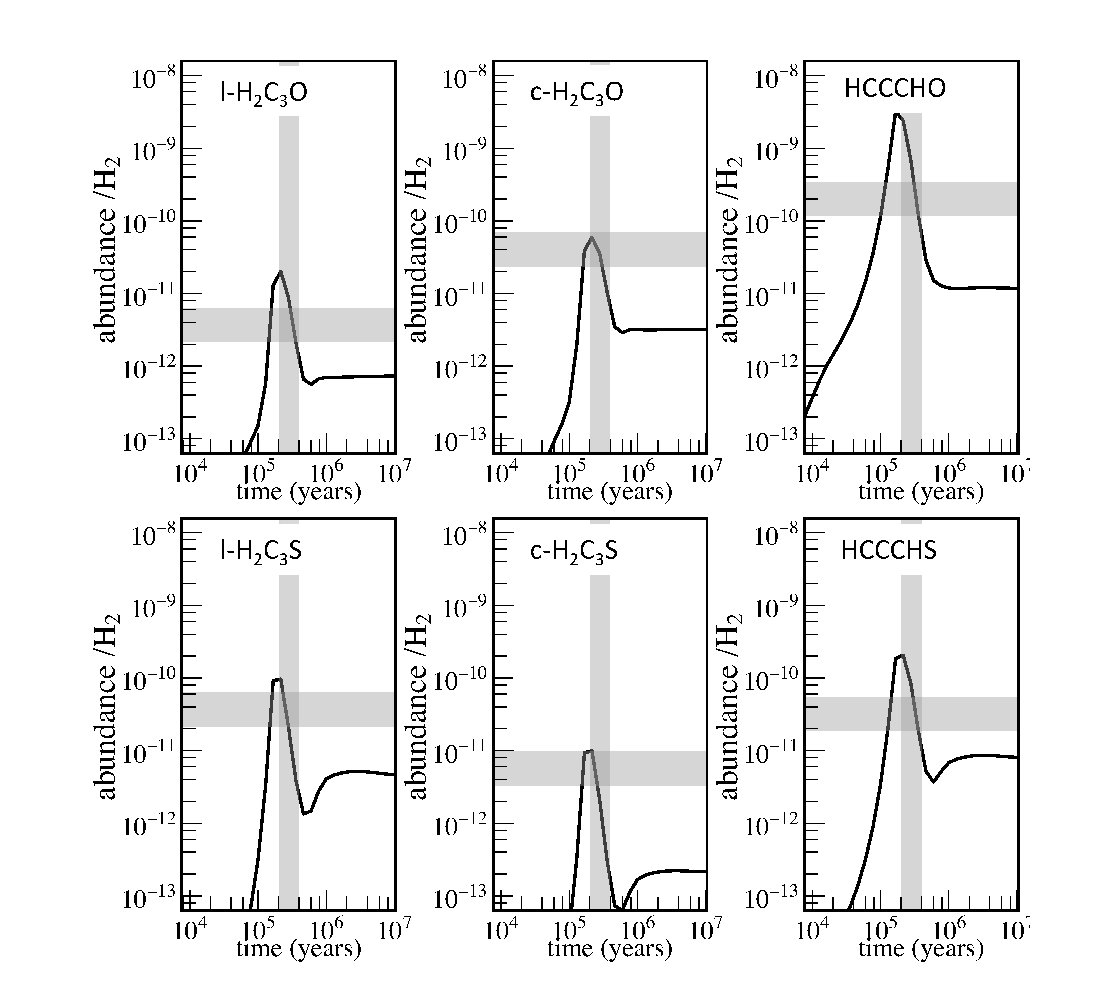}
\label{figure:l-H2C3O_model_Losion}
\end{SCfigure*}

Comparisons between our model and the various measurements for the H$_2$C$_3$O and H$_2$C$_3$S isomers is shown in Fig. \ref{figure:l-H2C3O_model_Losion}. 
There is good overall agreement given the many uncertainties. This is particularly critical for l-H$_2$C$_3$O, l-H$_2$C$_3$S, and c-H$_2$C$_3$S, where the DRs are potentially very significant but for which no experimental or theoretical data exist. The detection of l-H$_2$C$_3$O, which is the least abundant of the isomers despite being the most stable, completes the description of the H$_2$C$_3$O/H$_2$C$_3$S systems. This once again highlights that chemistry is controlled by kinetics rather than thermodynamics, although knowing the rates and branching ratios of the reactions involved in interstellar modelling is also crucial. It should be noted that the abundances of the H$_2$C$_3$O and H$_2$C$_3$S isomers are controlled by kinetics but with very different formation pathways for each isomer. Consequently, it is not a good example of the relative dipole principle (RDP) \citep[]{Shingledecker2020b} because the chemistries of the various isomers are too different. Indeed, even though their dipole moments are different, which affects the rate of destruction by ion-molecule ionisation and protonation reactions (with H$^{+}_{\mathrm{3}}$, HCO$^+$, and C$^+$), these differences in rates are relatively small compared with the large differences in their formation pathways. Furthermore, destruction reactions with neutrals such as atomic carbon and likely also with the OH radical are also important destruction reactions that are, a priori, similar for all isomers.

\section{Conclusions}
\label{section:summary}

We reported the first detection of propadienone (l-H$_2$C$_3$O) in the ISM, as well as the deuterated (c-HDC$_3$O) and $^{13}$C substitutions of cyclopropenone, through the highly sensitive QUIJOTE spectral line survey in TMC-1. Our analysis reveals that l-H$_2$C$_3$O is the least abundant isomer of this species, despite being the most stable. In particular, it is about one order of magnitude less abundant than the other two isomers (c-H$_2$C$_3$O and HCCCHO) and about twenty times less abundant than its sulphur analogue l-H$_2$C$_3$S. 
Chemical models also indicate that the chemistry responsible for the formation of the H$_2$C$_3$O and H$_2$C$_3$S isomers is very different, with the DR being the main production pathway for H$_2$C$_3$O, while H$_2$C$_3$S is mainly formed through gas phase ion-molecule chemistry.

\begin{acknowledgements}

This work was based on observations carried out with the Yebes 40m telescope (projects 19A003, 20A014, 20D023, 21A011, 21D005, and 23A024) and the IRAM 30m telescope. 
G.E., N.M., and B.T. acknowledge support from the Spanish grant PID2022-137980NB-I00, funded by MCIN/AEI/10.13039/501100011033/FEDER UE.
We acknowledge funding support from Spanish Ministerio de Ciencia e Innovación through grants PID2022-136525NA-I00 and PID2023-147545NB-I00. G. E. acknowledges the ERC project SUL4LIFE (grant agreement No101096293) from European Union. 
G. M. acknowledges the support of the grant RYC2022-035442-I funded by MICIU/AEI/10.13039/501100011033 and ESF+. G. M. also received support from project 20245AT016 (Proyectos Intramurales CSIC).

\end{acknowledgements}

\bibliographystyle{plain}
\bibliography{biblio}

\begin{appendix}

\section{Additional tables and figures}
\label{tables_figures}

This Appendix shows the line parameters (Table \ref{table:line_parameters}) obtained from Gaussian fits of the detected lines in TMC-1 (see Sect. \ref{section:results}), and the rates (Table \ref{table:rates}) for the DR of HCCCHSH$^+$ and C$_2$H$_3$CS$^+$ (see Sect. \ref{section:chemical_mechanisms_H2CCCO}). 

\begin{SCfigure*}[0.7][h]
\includegraphics[scale=0.63, angle=0]{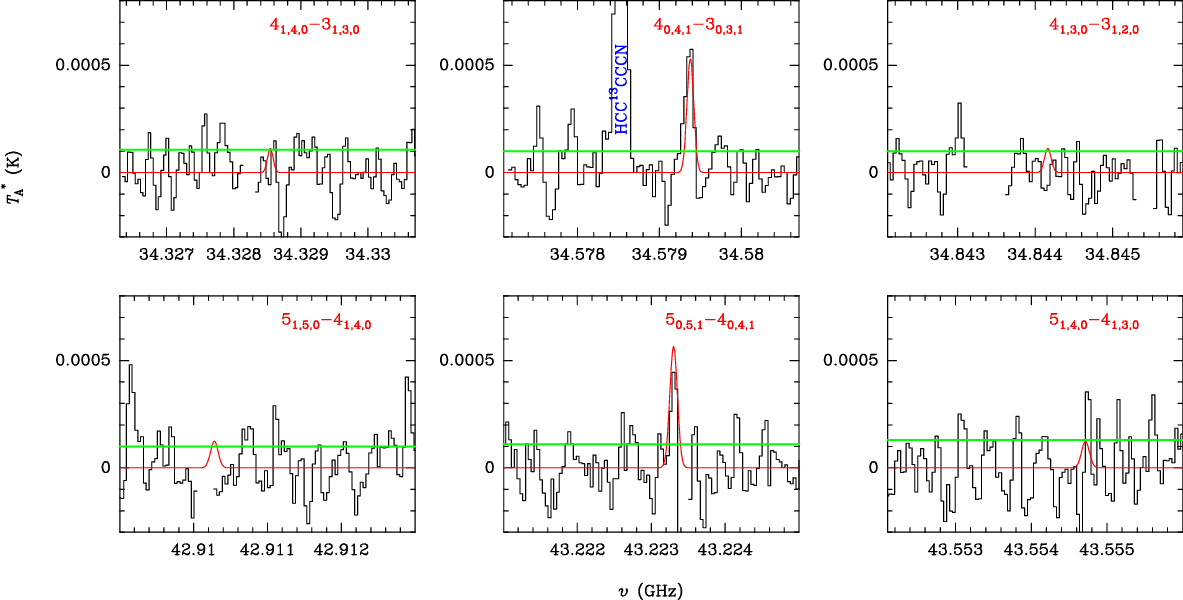} 
\hspace{1.5cm}
\\
\caption{Observed lines of l-H$_2$C$_3$O for $E$$_{\mathrm{upp}}$$<$15 K in TMC-1.}
\label{figure:o-H2CCCO_lines}
\end{SCfigure*}

\begin{figure*}
\vspace{0.3cm}
\hspace{0.9cm}
\includegraphics[scale=0.585, angle=0]{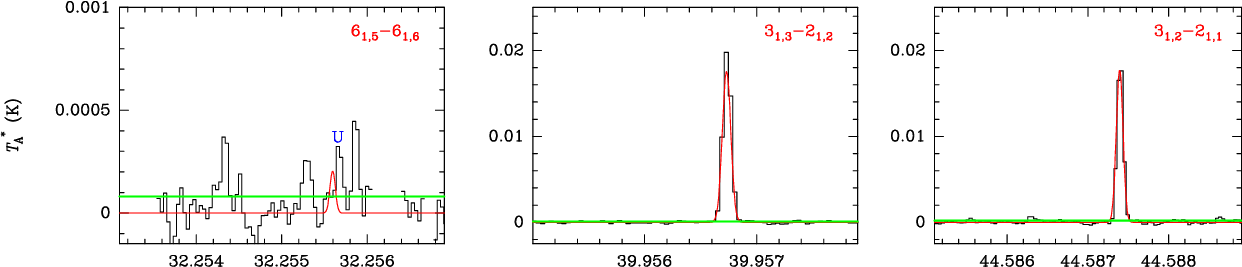} \vspace{0.5cm} \\
\vspace{0.7cm}
\hspace{0.9cm}
\includegraphics[scale=0.585, angle=0]{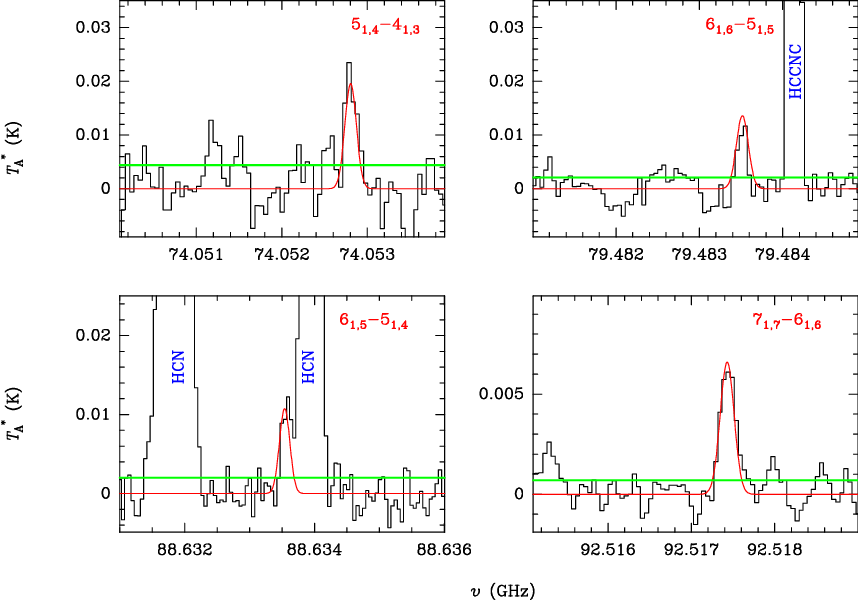}\\
\vspace{-1.0cm}
\hspace{0.9cm}
\caption{Observed lines of o-c-H$_2$C$_3$O in TMC-1 at 7 and 3 mm for $E$$_{\mathrm{upp}}$$\lesssim$17 K. }
\label{figure:o-c-H2C3O_lines}
\end{figure*}

\begin{SCfigure*}[0.4][h]
\includegraphics[scale=0.63, angle=0]{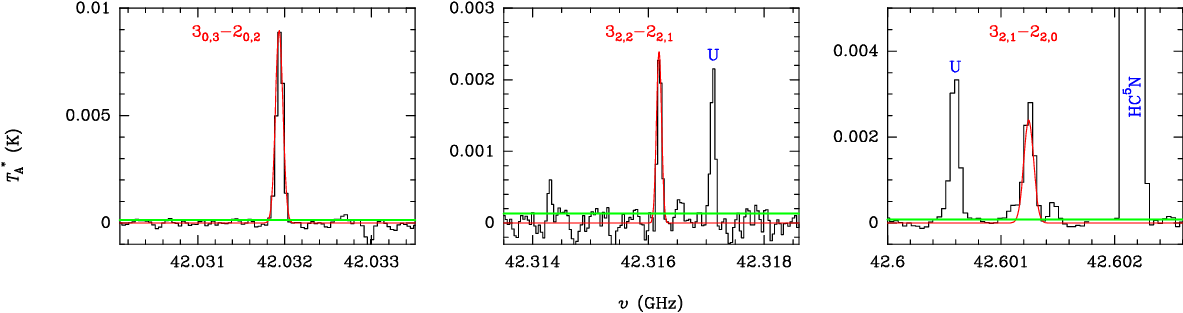} 
\hspace{1.5cm}
\\
\caption{Observed lines of p-c-H$_2$C$_3$O in TMC-1 in the 31.0-50.4 GHz range.}
\label{figure:p-c-H2C3O_lines}
\end{SCfigure*}

\begin{SCfigure*}[0.4][h]
\includegraphics[scale=0.63, angle=0]{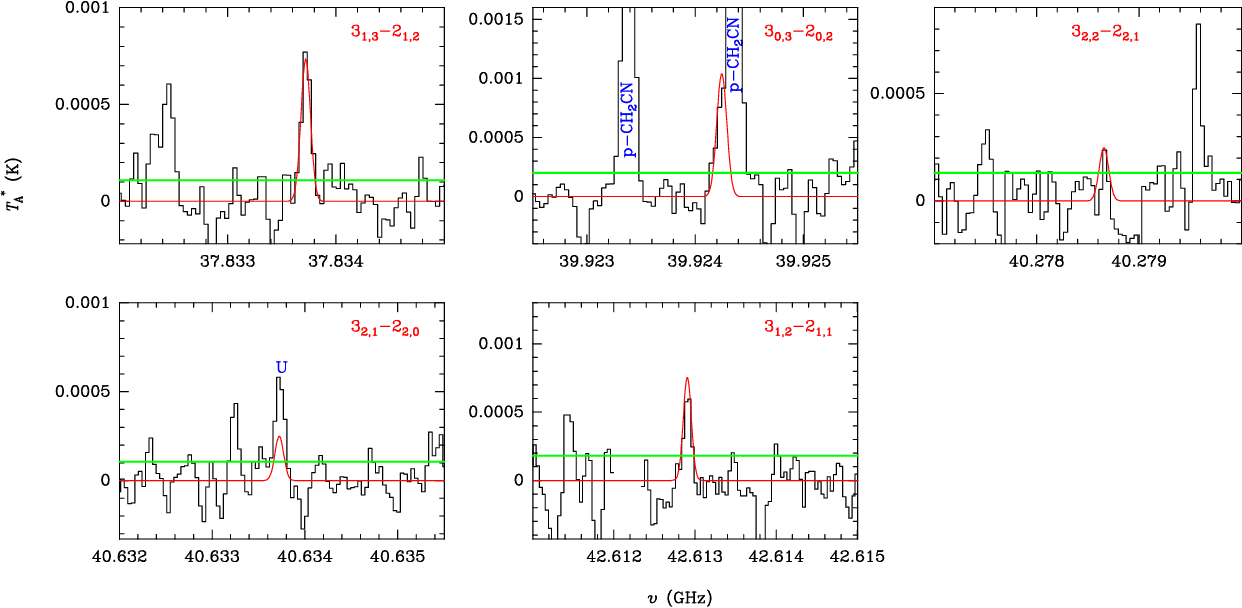} 
\hspace{1.5cm}
\\
\caption{Observed lines of c-HDC$_3$O for $E$$_{\mathrm{upp}}$$<$9 K and $A$$_{\mathrm{ul}}$$>$10$^{-6}$ s$^{-1}$ in TMC-1 in the 31.0-50.4 GHz range.}
\label{figure:c-HDC3O_lines}
\end{SCfigure*}

\begin{SCfigure*}[0.4][h]
\includegraphics[scale=0.63, angle=0]{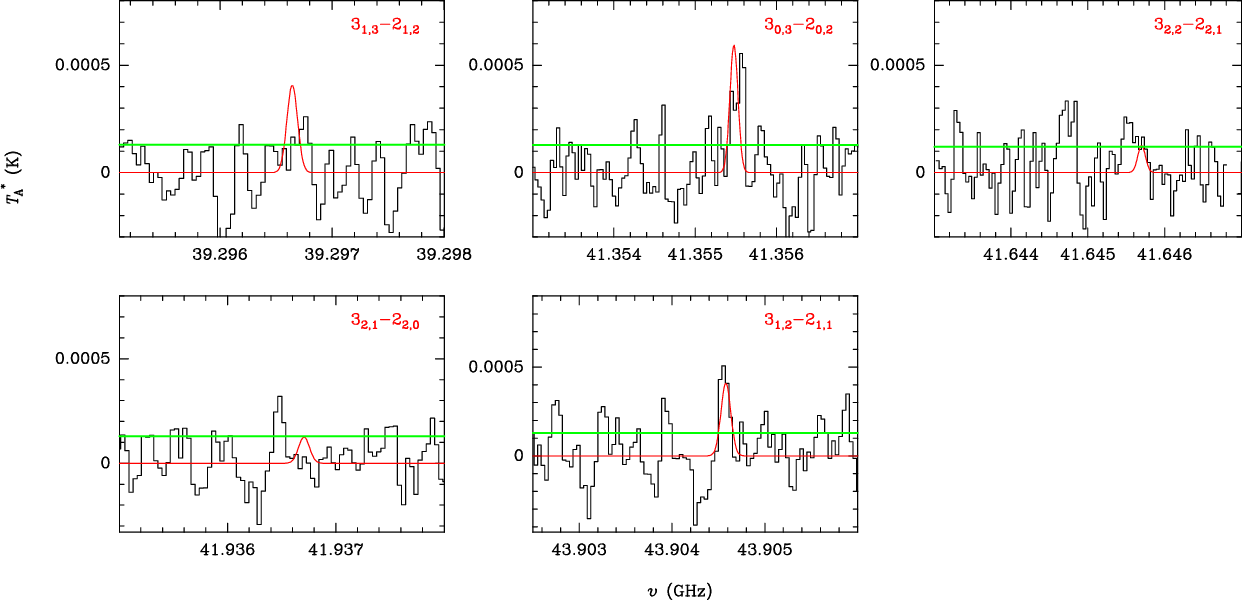} 
\hspace{1.5cm}
\\
\caption{Observed lines of c-H$_2$$^{13}$CCCO in TMC-1 in the 31.0-50.4 GHz range.}
\label{figure:c-H213CCCO_lines}
\end{SCfigure*}

\begin{SCfigure*}[0.9][h]
\includegraphics[scale=0.63, angle=0]{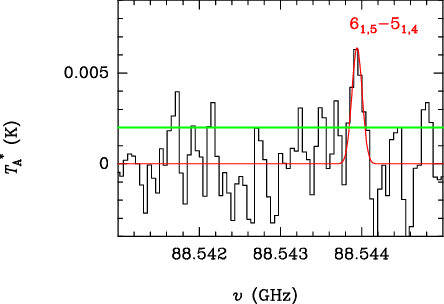} 
\hspace{1.5cm}
\\
\caption{Observed line of c-H$_2$CC$^{13}$CO in TMC-1 in the 31.0-50.4 GHz range.}
\label{figure:o-c-H2CC13CO_lines}
\end{SCfigure*}

\begin{SCfigure*}[0.4][h]
\includegraphics[scale=0.63, angle=0]{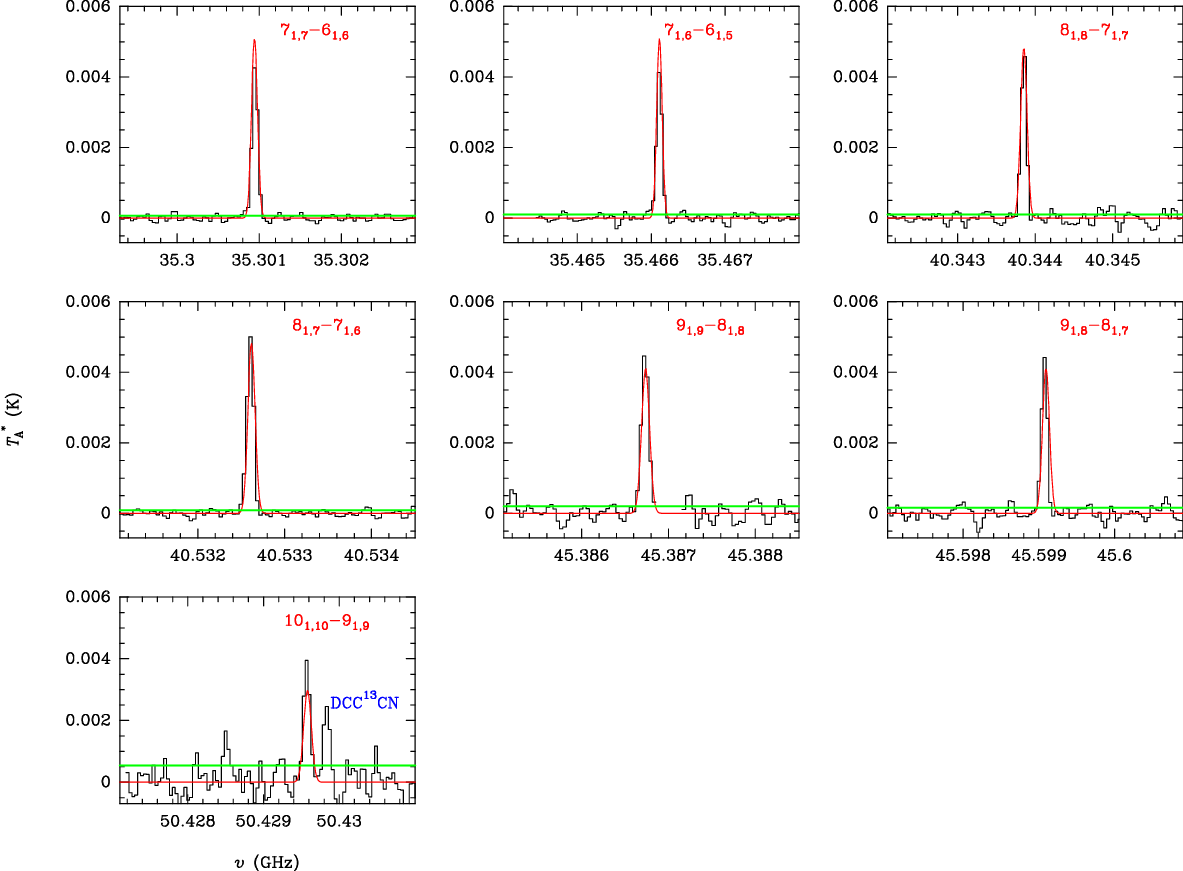} 
\hspace{1.5cm}
\\
\caption{Observed lines of o-H$_2$C$_3$S in TMC-1.}
\label{figure:o-H2CCCS_lines}
\end{SCfigure*}

\begin{SCfigure*}[0.4][h]
\includegraphics[scale=0.63, angle=0]{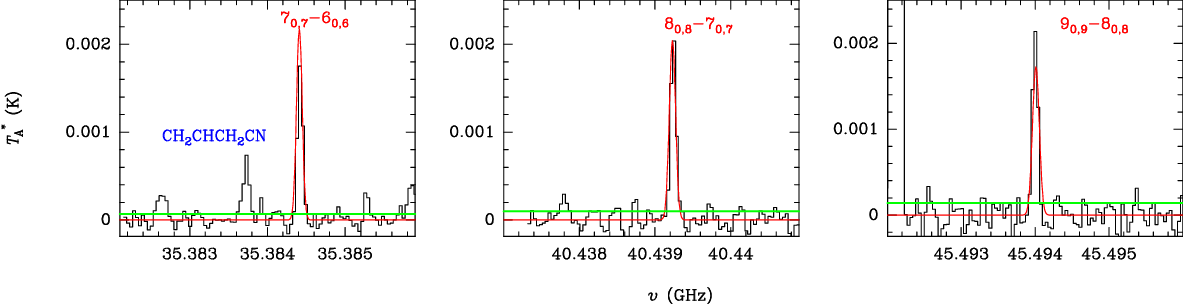} 
\hspace{1.5cm}
\\
\caption{Observed lines of p-H$_2$C$_3$S in TMC-1.}
\label{figure:p-H2CCCS_lines}
\end{SCfigure*}

\begin{SCfigure*}[0.4][h]
\includegraphics[scale=0.63, angle=0]{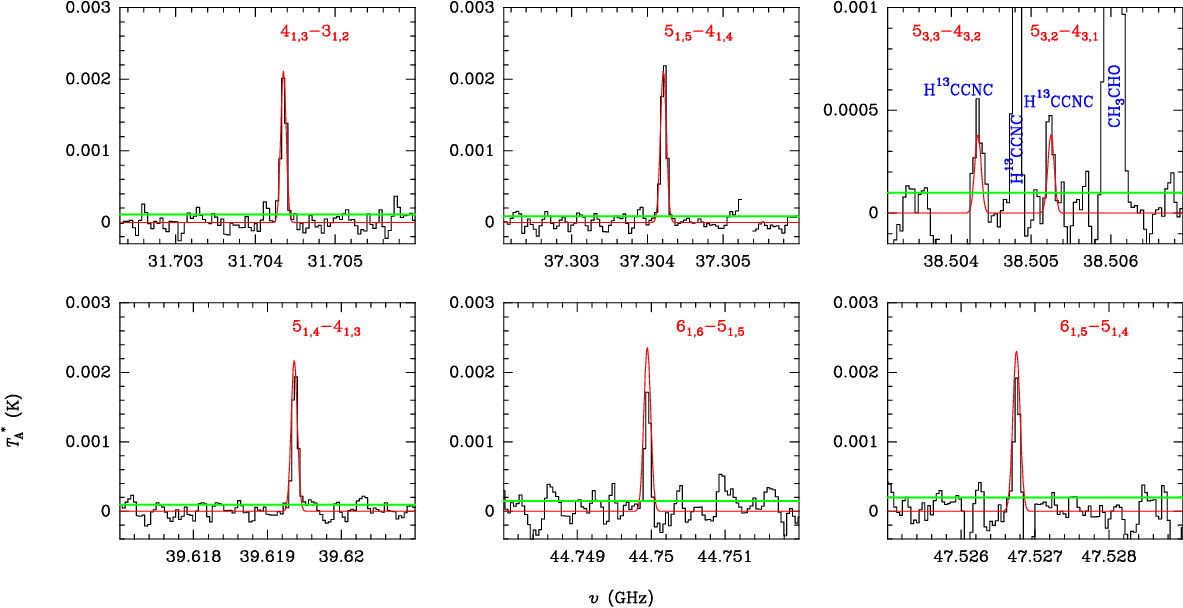} 
\hspace{1.5cm}
\\
\caption{Observed lines of o-c-H$_2$C$_3$S in TMC-1.}
\label{figure:o-c-H2CCCS_lines}
\end{SCfigure*}

\begin{SCfigure*}[0.4][h]
\includegraphics[scale=0.63, angle=0]{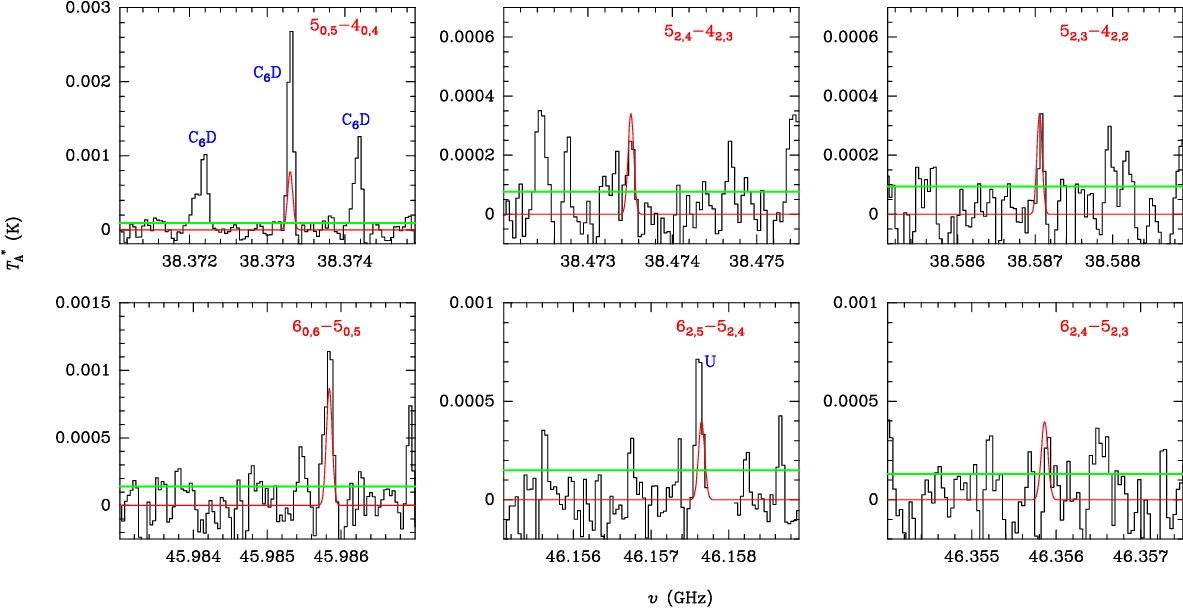} 
\hspace{1.5cm}
\\
\caption{Observed lines of p-c-H$_2$C$_3$S in TMC-1 for $E$$_{\mathrm{upp}}$$<$14 K.}
\label{figure:p-c-H2CCCS_lines}
\end{SCfigure*}

\begin{figure*}
\vspace{0.3cm}
\hspace{0.9cm}
\includegraphics[scale=0.585, angle=0]{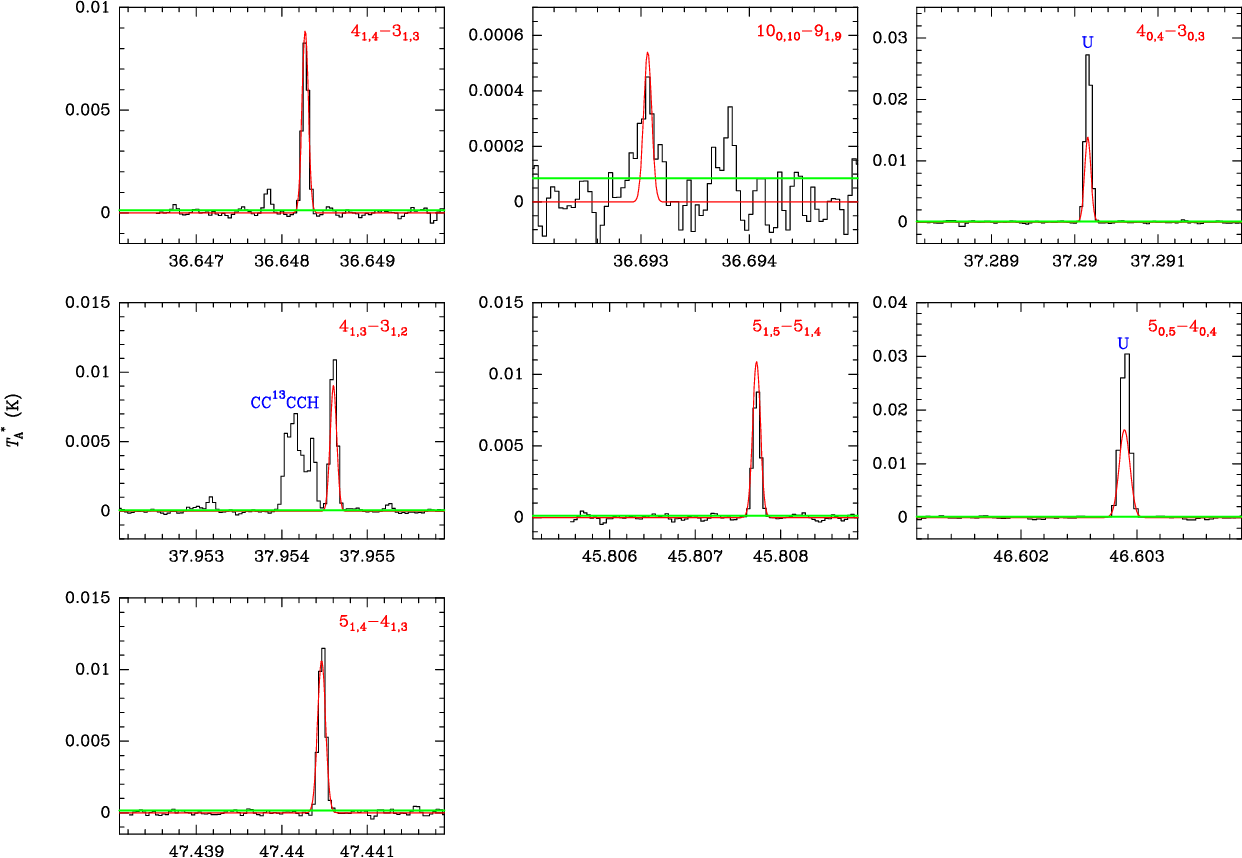} \vspace{0.5cm} \\
\vspace{0.7cm}
\hspace{0.9cm}
\includegraphics[scale=0.585, angle=0]{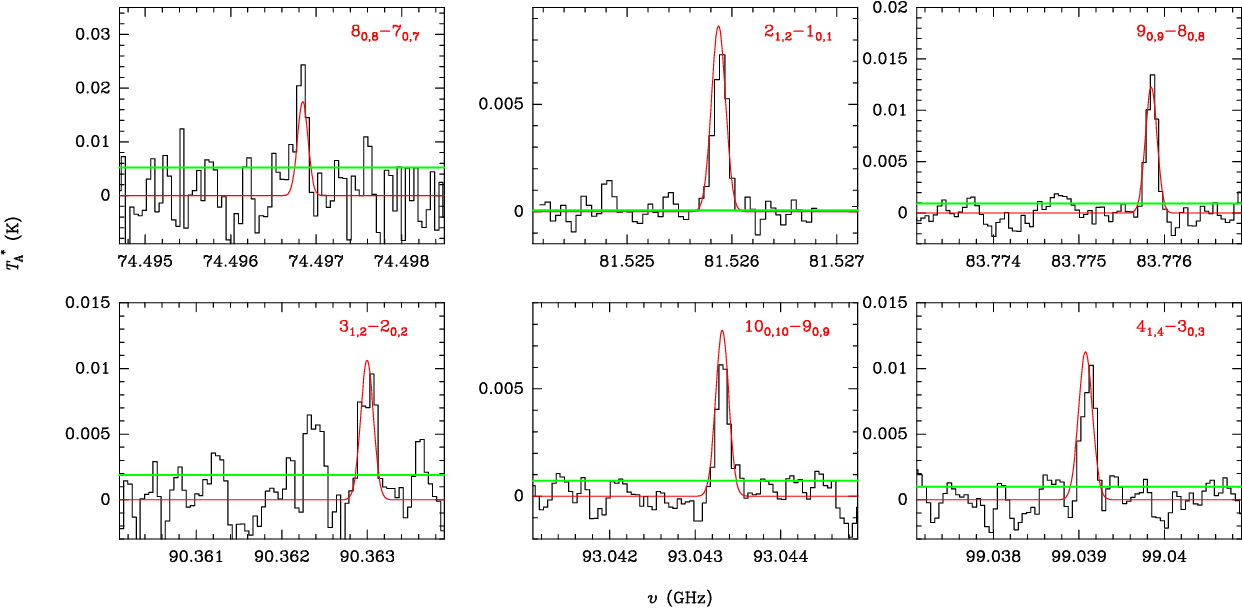}\\
\vspace{-1.0cm}
\hspace{0.9cm}
\caption{Observed lines of HCCCHO in TMC-1 at 7 and 3 mm.}
\label{figure:HCCCHO_lines}
\end{figure*}

\begin{figure*}
\includegraphics[scale=0.63, angle=0]{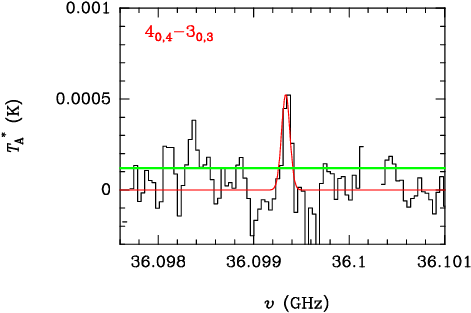} 
\hspace{1.5cm}
\\
\caption{Observed line of H$^{13}$CCCHO in TMC-1.}
\label{figure:H13CCCHO_lines}
\end{figure*}

\begin{figure*}
\includegraphics[scale=0.63, angle=0]{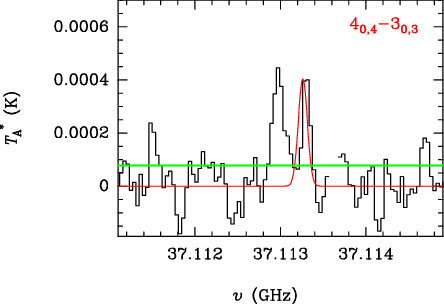} 
\hspace{1.5cm}
\\
\caption{Observed line of HC$^{13}$CCHO in TMC-1.}
\label{figure:HC13CCHO_lines}
\end{figure*}

\begin{figure*}
\centering
\includegraphics[scale=0.39, angle=0]{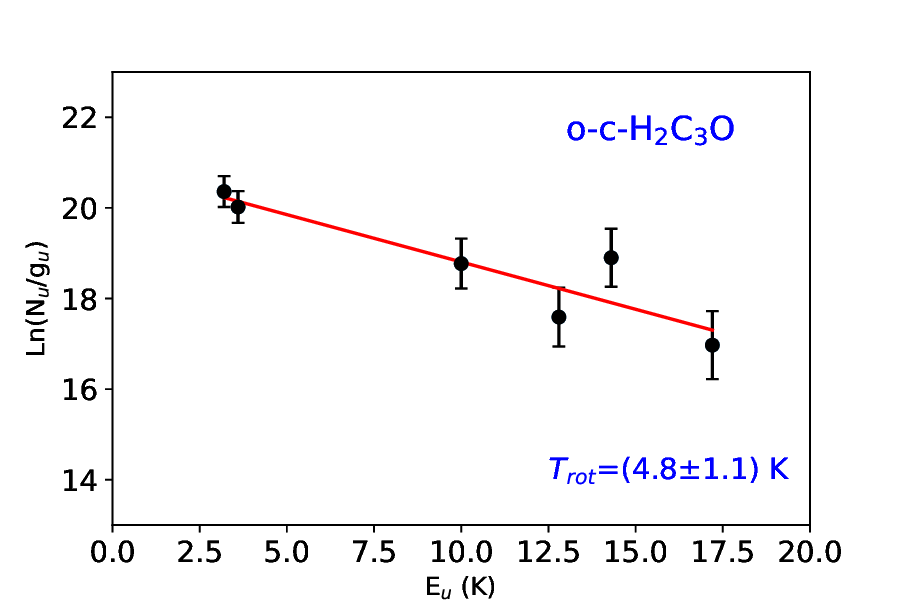}  
\includegraphics[scale=0.39, angle=0]{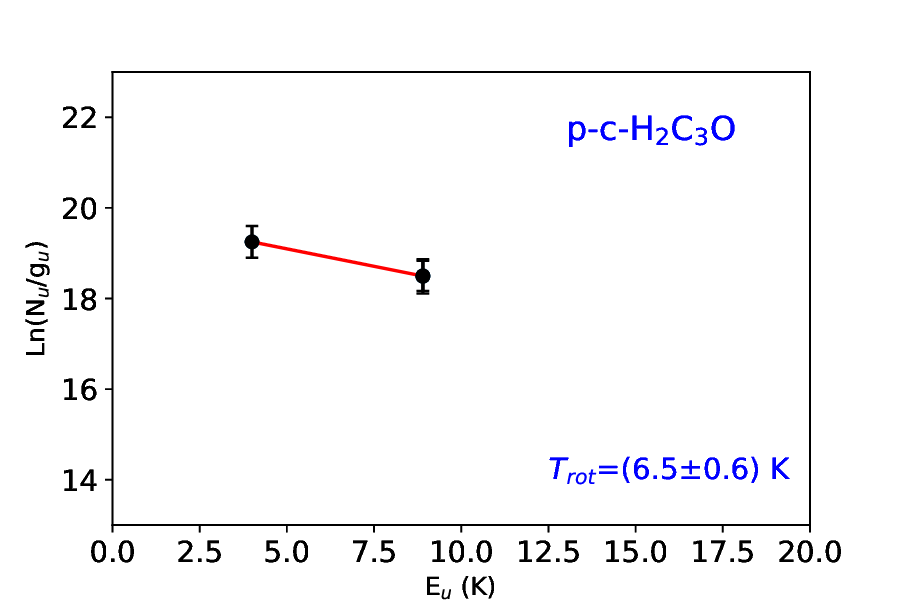} 
\includegraphics[scale=0.39, angle=0]{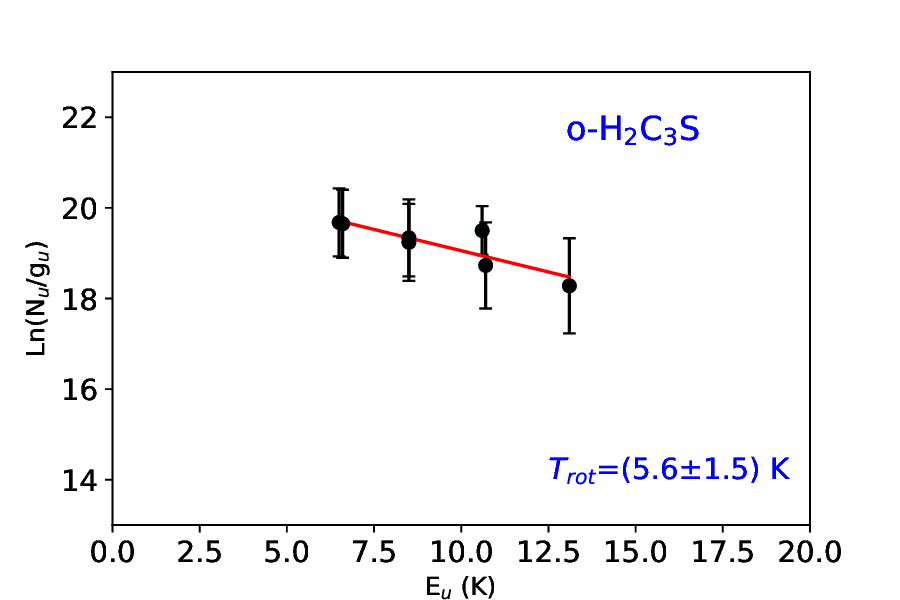} 

\includegraphics[scale=0.39, angle=0]{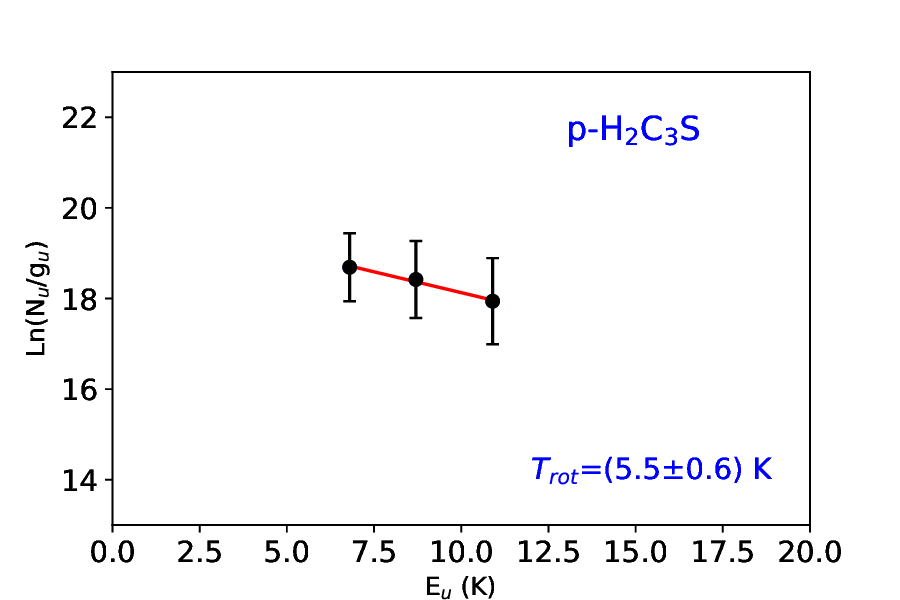}  
\includegraphics[scale=0.39, angle=0]{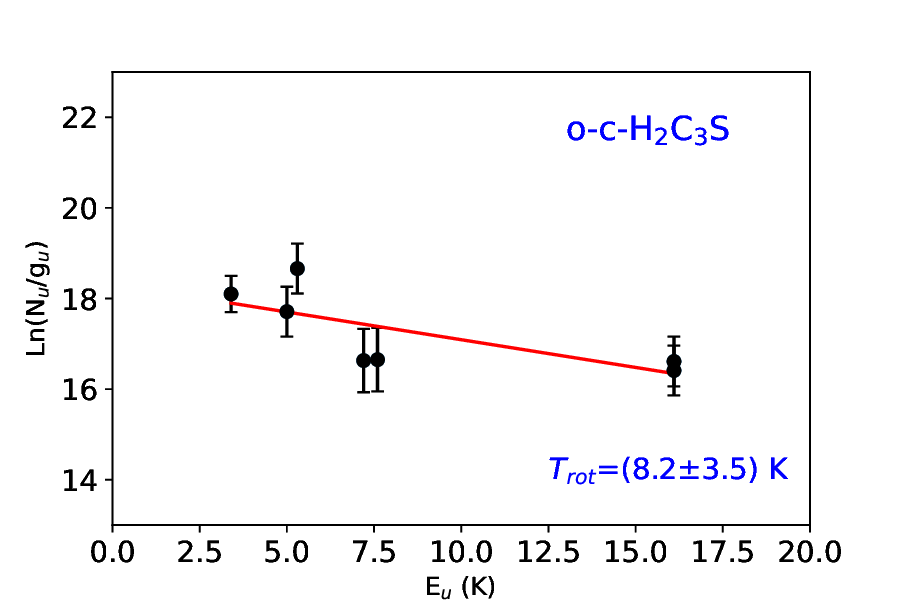}  
\includegraphics[scale=0.39, angle=0]{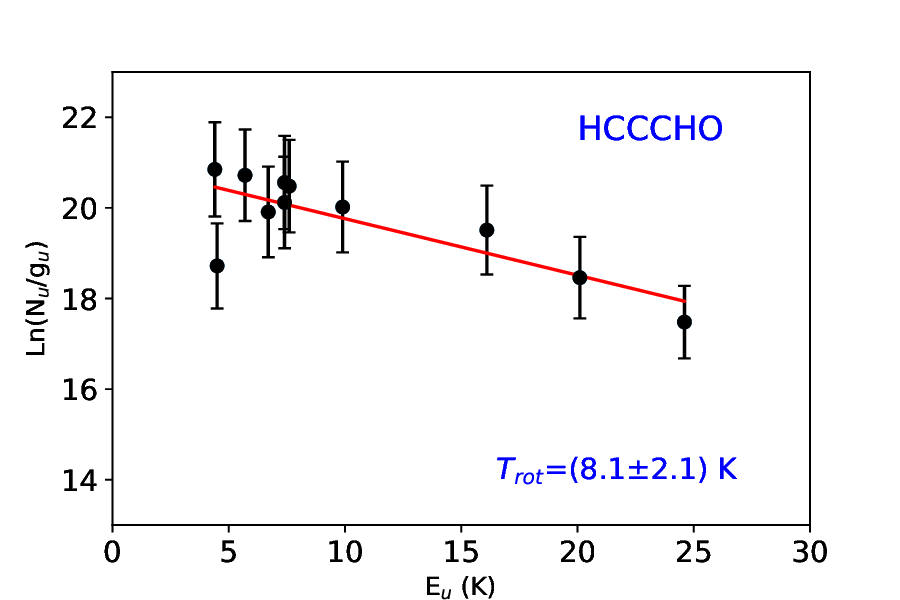} 
  
\hspace{1cm}
\\
\caption{Rotational diagrams for o-c-H$_2$C$_3$O, p-c-H$_2$C$_3$O, o-H$_2$C$_3$S, p-H$_2$C$_3$S, o-c-H$_2$C$_3$S, and HCCCHO. The fitted values of the rotational temperature, $T$$_{\mathrm{rot}}$, and its respective uncertainty are also indicated for each molecule.}
\label{figure:RD}
\end{figure*}

Figures \ref{figure:o-H2CCCO_lines}-\ref{figure:HC13CCHO_lines} show the observed lines listed in Table \ref{table:line_parameters}, with their quantum numbers indicated in each panel. The red line shows the LTE synthetic spectrum from a fit to the observed line profiles, and the horizontal green line indicates the 1$\sigma$ noise level. Figure \ref{figure:RD} shows the rotational diagrams derived in Sect. \ref{section:rotational_diagrams}.

\onecolumn

\begin{longtable}{lrlrlrlrl}
\caption{Line parameters obtained from Gaussian fits of the detected lines in TMC-1\label{table:line_parameters}.}\\
\hline\hline 
         & Transition                   & Rest          &  $E$$_{\mathrm{up}}$ & $A$$_{\mathrm{ul}}$ &  v$_{\mathrm{LSR}}$ & $\Delta$v     & $\int$$T$$^{\star}_{\mathrm{A}}$dv  &  $T$$^{\star}_{\mathrm{A}}$ \\
Species  & $J_{K_u,k_u}-J'_{K'_l,k'_l}$ & Frequency     &  (K)                 & (s$^{-1}$)          & (km s$^{-1}$)       & (km s$^{-1}$) & (mK km s$^{-1}$)                    &  (mK)                      \\
         &                              & (MHz)         &                      &                     &                     &               &                                     &                            \\
\hline
\endfirsthead
\caption{continued.}\\
\hline\hline
         & Transition                   & Rest          &  $E$$_{\mathrm{up}}$ & $A$$_{\mathrm{ul}}$ &  v$_{\mathrm{LSR}}$ & $\Delta$v     & $\int$$T$$^{\star}_{\mathrm{A}}$dv  &  $T$$^{\star}_{\mathrm{A}}$ \\
Species  & $J_{K_u,k_u}-J'_{K'_l,k'_l}$ & Frequency     &  (K)                 & (s$^{-1}$)          & (km s$^{-1}$)       & (km s$^{-1}$) & (mK km s$^{-1}$)                    &  (mK)                      \\
         &                              & (MHz)         &                      &                     &                     &               &                                     &                            \\
\hline
\endhead
\hline
\endfoot
l-H$_2$C$_3$O$^{a}$ & 4$_{0}$$_{,}$$_{4}$-3$_{0}$$_{,}$$_{3}$ &  34579.37  & 4.3  & 9.9$\times$e(-7) & 5.96$\pm$0.24   & 1.10$\pm$0.09  & 0.66$\pm$0.20  & 0.6  \\
l-H$_2$C$_3$O$^{b}$ & 5$_{0}$$_{,}$$_{5}$-4$_{0}$$_{,}$$_{4}$ &  43223.30  & 6.4  & 1.9$\times$e(-6) & 5.86$\pm$0.09   & 0.54$\pm$0.09  & 0.25$\pm$0.08  & 0.4  \\
\hline
o-c-H$_2$C$_3$O$^{c}$  & 6$_{1}$$_{,}$$_{5}$-6$_{1}$$_{,}$$_{6}$ &  32255.60  & 14.3 & 1.8$\times$e(-7) & -   & -  & -  & -  \\
o-c-H$_2$C$_3$O        & 3$_{1}$$_{,}$$_{3}$-2$_{1}$$_{,}$$_{2}$ &  39956.73  & 3.2  & 5.4$\times$e(-6) & 5.78$\pm$0.03   & 0.70$\pm$0.04  & 14.70$\pm$0.5  & 19.7  \\
o-c-H$_2$C$_3$O        & 3$_{1}$$_{,}$$_{2}$-2$_{1}$$_{,}$$_{1}$ &  44587.39  & 3.6  & 7.5$\times$e(-6) & 5.79$\pm$0.03   & 0.73$\pm$0.07  & 13.6$\pm$0.2   & 17.5 \\
o-c-H$_2$C$_3$O        & 5$_{1}$$_{,}$$_{4}$-4$_{1}$$_{,}$$_{3}$ &  74052.80  & 10.0 & 3.9$\times$e(-5) & 5.82$\pm$0.05   & 0.58$\pm$0.15  & 14.58$\pm$0.3  & 23.6 \\
o-c-H$_2$C$_3$O        & 6$_{1}$$_{,}$$_{6}$-5$_{1}$$_{,}$$_{5}$ &  79483.51  & 12.8 & 5.0$\times$e(-5) & 5.82$\pm$0.04   & 0.49$\pm$0.08  & 6.20$\pm$0.5   & 11.0 \\
o-c-H$_2$C$_3$O        & 6$_{1}$$_{,}$$_{5}$-5$_{1}$$_{,}$$_{4}$ &  88633.53  & 14.3 & 6.9$\times$e(-5) & 5.80$\pm$0.04   & 0.50$\pm$0.11  & 6.90$\pm$0.5   & 13.0 \\
o-c-H$_2$C$_3$O        & 7$_{1}$$_{,}$$_{7}$-6$_{1}$$_{,}$$_{6}$ &  92517.42  & 17.2 & 8.1$\times$e(-5) & 5.78$\pm$0.03   & 0.79$\pm$0.06  & 5.10$\pm$0.6   & 6.1 \\
\hline
p-c-H$_2$C$_3$O        & 3$_{0}$$_{,}$$_{3}$-2$_{0}$$_{,}$$_{2}$ &  42031.93  & 4.0  & 7.1$\times$e(-6) & 5.78$\pm$0.06   & 0.65$\pm$0.02  & 6.34$\pm$0.07  & 9.0  \\
p-c-H$_2$C$_3$O        & 3$_{2}$$_{,}$$_{2}$-2$_{2}$$_{,}$$_{1}$ &  42316.18  & 8.9  & 4.0$\times$e(-6) & 5.77$\pm$0.03   & 0.69$\pm$0.07  & 1.66$\pm$0.09  & 2.2  \\
p-c-H$_2$C$_3$O        & 3$_{2}$$_{,}$$_{1}$-2$_{2}$$_{,}$$_{0}$ &  42601.24  & 8.9  & 4.1$\times$e(-6) & 5.82$\pm$0.03   & 0.74$\pm$0.09  & 2.23$\pm$0.04  & 2.8  \\
\hline
c-HDC$_3$O         & 3$_{1}$$_{,}$$_{3}$-2$_{1}$$_{,}$$_{2}$ &  37833.72  & 4.7  & 4.6$\times$e(-6) & 5.79$\pm$0.06   & 0.82$\pm$0.15  & 0.67$\pm$0.15  & 0.7  \\
c-HDC$_3$O$^{d}$   & 3$_{0}$$_{,}$$_{3}$-2$_{0}$$_{,}$$_{2}$ &  39924.24  & 3.8  & 6.1$\times$e(-6) & -   & -  & -  & -  \\
c-HDC$_3$O$^{*}$   & 3$_{2}$$_{,}$$_{2}$-2$_{2}$$_{,}$$_{1}$ &  40278.65  & 8.0  & 3.5$\times$e(-6) & -   & -  & -  & -  \\
c-HDC$_3$O$^{c}$   & 3$_{2}$$_{,}$$_{1}$-2$_{2}$$_{,}$$_{0}$ &  40633.72  & 8.0  & 3.5$\times$e(-6) & -   & -  & -  & -  \\
c-HDC$_3$O         & 3$_{1}$$_{,}$$_{2}$-2$_{1}$$_{,}$$_{1}$ &  42612.90  & 5.1  & 6.6$\times$e(-6) & 5.80$\pm$0.09   & 0.90$\pm$0.22  & 0.52$\pm$0.28  & 0.6  \\
\hline
c-H$_2$$^{13}$CCCO$^{*}$   & 3$_{1}$$_{,}$$_{3}$-2$_{1}$$_{,}$$_{2}$ &  39296.64  & 4.9  & 5.2$\times$e(-6) & -   & -  & -  & -  \\
c-H$_2$$^{13}$CCCO         & 3$_{0}$$_{,}$$_{3}$-2$_{0}$$_{,}$$_{2}$ &  41355.47  & 4.0  & 6.8$\times$e(-6) & 5.31$\pm$0.12   & 0.85$\pm$0.25  & 0.49$\pm$0.17  & 0.5  \\
c-H$_2$$^{13}$CCCO$^{*}$   & 3$_{2}$$_{,}$$_{2}$-2$_{2}$$_{,}$$_{1}$ &  41645.70  & 8.6  & 3.8$\times$e(-6) & -   & -  & -  & -  \\
c-H$_2$$^{13}$CCCO$^{*}$   & 3$_{2}$$_{,}$$_{1}$-2$_{2}$$_{,}$$_{0}$ &  41936.71  & 8.7  & 3.9$\times$e(-6) & -   & -  & -  & -  \\
c-H$_2$$^{13}$CCCO         & 3$_{1}$$_{,}$$_{2}$-2$_{1}$$_{,}$$_{1}$ &  43904.55  & 5.4  & 7.2$\times$e(-6) & 5.85$\pm$0.10   & 0.81$\pm$0.15  & 0.44$\pm$0.18  & 0.5  \\
\hline
c-H$_2$CC$^{13}$CO & 6$_{1}$$_{,}$$_{5}$-5$_{1}$$_{,}$$_{4}$ &  88543.94  & 14.3 & 6.9$\times$e(-5) & 5.84$\pm$0.07   & 0.52$\pm$0.14  & 3.51$\pm$1.0  & 6.2  \\
\hline
o-H$_2$C$_3$S       & 7$_{1}$$_{,}$$_{7}$-6$_{1}$$_{,}$$_{6}$ &  35300.94  & 6.5 & 9.9$\times$e(-7) & 5.79$\pm$0.06   & 0.76$\pm$0.01  & 3.44$\pm$0.4  & 4.2  \\
o-H$_2$C$_3$S       & 7$_{1}$$_{,}$$_{6}$-6$_{1}$$_{,}$$_{5}$ &  35466.11  & 6.6 & 1.0$\times$e(-6) & 5.83$\pm$0.08   & 0.77$\pm$0.02  & 3.37$\pm$0.2  & 4.1  \\
o-H$_2$C$_3$S       & 8$_{1}$$_{,}$$_{8}$-7$_{1}$$_{,}$$_{7}$ &  40343.85  & 8.5 & 1.5$\times$e(-6) & 5.85$\pm$0.09   & 0.68$\pm$0.02  & 3.33$\pm$0.2  & 4.5  \\
o-H$_2$C$_3$S       & 8$_{1}$$_{,}$$_{7}$-7$_{1}$$_{,}$$_{6}$ &  40532.61  & 8.5 & 1.5$\times$e(-6) & 5.92$\pm$0.06   & 0.69$\pm$0.03  & 3.69$\pm$0.5  & 5.0  \\
o-H$_2$C$_3$S       & 9$_{1}$$_{,}$$_{9}$-8$_{1}$$_{,}$$_{8}$ &  45386.72  & 10.6 & 2.2$\times$e(-6) & 5.87$\pm$0.04   & 0.75$\pm$0.03  & 3.52$\pm$0.7  & 4.1  \\
o-H$_2$C$_3$S       & 9$_{1}$$_{,}$$_{8}$-8$_{1}$$_{,}$$_{7}$ &  45599.08  & 10.7 & 2.2$\times$e(-6) & 5.96$\pm$0.02   & 0.60$\pm$0.02  & 2.88$\pm$0.3  & 4.4  \\
o-H$_2$C$_3$S       & 10$_{1}$$_{,}$$_{10}$-9$_{1}$$_{,}$$_{9}$ & 50429.58  & 13.1 & 3.0$\times$e(-6) & 5.89$\pm$0.03   & 0.61$\pm$0.09  & 2.60$\pm$0.5  & 4.0  \\
\hline
p-H$_2$C$_3$S       & 7$_{0}$$_{,}$$_{7}$-6$_{0}$$_{,}$$_{6}$ & 35384.41  & 6.8 & 1.0$\times$e(-6) & 5.74$\pm$0.02   & 0.69$\pm$0.05  & 1.31$\pm$0.08  & 1.7  \\
p-H$_2$C$_3$S       & 8$_{0}$$_{,}$$_{8}$-7$_{0}$$_{,}$$_{7}$ & 40439.22  & 8.7 & 1.5$\times$e(-6) & 5.74$\pm$0.02   & 0.65$\pm$0.06  & 1.45$\pm$0.07  & 2.1  \\
p-H$_2$C$_3$S       & 9$_{0}$$_{,}$$_{9}$-8$_{0}$$_{,}$$_{8}$ & 45944.00 & 10.9 & 2.2$\times$e(-6) & 5.86$\pm$0.05   & 0.58$\pm$0.09  & 1.32$\pm$0.08  & 2.1  \\
\hline
o-c-H$_2$C$_3$S     & 4$_{1}$$_{,}$$_{3}$-3$_{1}$$_{,}$$_{2}$ & 31704.35  & 3.4 & 3.2$\times$e(-6) & 5.80$\pm$0.02   & 0.80$\pm$0.05  & 1.72$\pm$0.05  & 2.0  \\
o-c-H$_2$C$_3$S     & 5$_{1}$$_{,}$$_{5}$-4$_{1}$$_{,}$$_{4}$ & 37304.21  & 5.0 & 5.5$\times$e(-6) & 5.77$\pm$0.01   & 0.75$\pm$0.03  & 1.76$\pm$0.05  & 2.2  \\
o-c-H$_2$C$_3$S$^{e}$     & 5$_{3}$$_{,}$$_{3}$-4$_{3}$$_{,}$$_{2}$ & 38504.33  & 16.1 & 4.0$\times$e(-6) & 5.80$\pm$0.01   & 0.80$\pm$0.04  & 0.46$\pm$0.03  & 0.5  \\
o-c-H$_2$C$_3$S$^{e}$     & 5$_{3}$$_{,}$$_{2}$-4$_{3}$$_{,}$$_{1}$ & 38505.25  & 16.1 & 4.0$\times$e(-6) & 5.92$\pm$0.04   & 0.69$\pm$0.04  & 0.37$\pm$0.03  & 0.5  \\
o-c-H$_2$C$_3$S     & 5$_{1}$$_{,}$$_{4}$-4$_{1}$$_{,}$$_{3}$ & 39619.37  & 5.3  & 6.5$\times$e(-6) & 5.76$\pm$0.02   & 0.75$\pm$0.04  & 1.54$\pm$0.06  & 1.9  \\
o-c-H$_2$C$_3$S     & 6$_{1}$$_{,}$$_{6}$-5$_{1}$$_{,}$$_{5}$ & 44749.94  & 7.2  & 9.7$\times$e(-6) & 5.83$\pm$0.03   & 0.62$\pm$0.05  & 1.12$\pm$0.05  & 1.7  \\
o-c-H$_2$C$_3$S     & 6$_{1}$$_{,}$$_{5}$-5$_{1}$$_{,}$$_{4}$ & 47526.74  & 7.6  & 1.1$\times$e(-5) & 5.80$\pm$0.03   & 0.55$\pm$0.07  & 1.16$\pm$0.04  & 1.9  \\
\hline
p-c-H$_2$C$_3$S$^{f}$ & 5$_{0}$$_{,}$$_{5}$-4$_{0}$$_{,}$$_{4}$ & 38373.30  & 5.5  & 6.2$\times$e(-6) & -   & - & -  & -  \\
p-c-H$_2$C$_3$S       & 5$_{2}$$_{,}$$_{4}$-4$_{2}$$_{,}$$_{3}$ & 38473.50  & 11.0 & 5.2$\times$e(-6) & 5.85$\pm$0.10   & 0.82$\pm$0.30  & 0.22$\pm$0.06  & 0.3  \\
p-c-H$_2$C$_3$S       & 5$_{2}$$_{,}$$_{3}$-4$_{2}$$_{,}$$_{2}$ & 38587.05  & 11.0 & 5.3$\times$e(-6) & 5.64$\pm$0.08   & 0.52$\pm$0.21  & 0.19$\pm$0.04  & 0.3  \\
p-c-H$_2$C$_3$S       & 6$_{0}$$_{,}$$_{6}$-5$_{0}$$_{,}$$_{5}$ & 45985.84  & 7.7  & 1.1$\times$e(-5) & 5.87$\pm$0.05   & 0.82$\pm$0.10  & 1.02$\pm$0.03  & 1.5  \\
p-c-H$_2$C$_3$S$^{c}$ & 6$_{2}$$_{,}$$_{5}$-5$_{2}$$_{,}$$_{4}$ & 46157.64  & 13.2 & 9.8$\times$e(-6) & -   & -  & -  & -  \\
p-c-H$_2$C$_3$S$^{*}$ & 6$_{2}$$_{,}$$_{4}$-5$_{2}$$_{,}$$_{3}$ & 46355.86  & 13.2 & 9.9$\times$e(-6) & -   & -  & -  & -  \\
\hline
HCCCHO          & 4$_{1}$$_{,}$$_{4}$-3$_{1}$$_{,}$$_{3}$   & 36648.27  & 7.4  & 1.3$\times$e(-6) & 5.80$\pm$0.08   & 0.73$\pm$0.02  & 6.4$\pm$0.3  & 8.3  \\
HCCCHO          & 10$_{0}$$_{,}$$_{10}$-9$_{1}$$_{,}$$_{9}$ & 36693.06  & 4.5  & 2.8$\times$e(-7) & 5.93$\pm$0.10   & 1.10$\pm$0.22  & 0.52$\pm$0.02 & 0.4  \\
HCCCHO$^{c}$    & 4$_{0}$$_{,}$$_{4}$-3$_{0}$$_{,}$$_{3}$   & 37290.15  & 4.5  & 1.4$\times$e(-6) & -   & - & -  & -  \\
HCCCHO          & 4$_{1}$$_{,}$$_{3}$-3$_{1}$$_{,}$$_{2}$   & 37954.60  & 7.6  & 1.4$\times$e(-6) & 5.79$\pm$0.04   & 0.81$\pm$0.08  & 9.4$\pm$0.1   & 10.9  \\
HCCCHO          & 5$_{1}$$_{,}$$_{5}$-4$_{1}$$_{,}$$_{4}$   & 45807.71  & 9.6  & 2.7$\times$e(-6) & 5.81$\pm$0.05   & 0.71$\pm$0.01  & 6.6$\pm$0.3   & 8.8  \\
HCCCHO$^{c}$    & 5$_{0}$$_{,}$$_{5}$-4$_{0}$$_{,}$$_{4}$   & 46602.89  & 6.7  & 2.9$\times$e(-6) & -   & -  & -  & -  \\
HCCCHO          & 5$_{1}$$_{,}$$_{4}$-4$_{1}$$_{,}$$_{3}$   & 47440.46  & 9.9  & 3.0$\times$e(-6) & 5.79$\pm$0.04   & 0.67$\pm$0.01  & 8.1$\pm$0.4   & 11.4  \\
HCCCHO          & 8$_{0}$$_{,}$$_{8}$-7$_{0}$$_{,}$$_{7}$   & 74496.68  & 16.1 & 1.2$\times$e(-5) & 5.89$\pm$0.06   & 0.58$\pm$0.09  & 14.9$\pm$0.7  & 24.0  \\
HCCCHO          & 2$_{1}$$_{,}$$_{2}$-1$_{0}$$_{,}$$_{1}$   & 81525.87  & 4.4  & 4.0$\times$e(-6) & 5.75$\pm$0.02   & 0.58$\pm$0.05  & 4.6$\pm$0.3   & 7.4  \\
HCCCHO          & 9$_{0}$$_{,}$$_{9}$-8$_{0}$$_{,}$$_{8}$   & 83775.84  & 20.1 & 1.8$\times$e(-5) & 5.80$\pm$0.02   & 0.49$\pm$0.03  & 7.1$\pm$0.5   & 13.5  \\
HCCCHO          & 3$_{1}$$_{,}$$_{3}$-2$_{0}$$_{,}$$_{2}$   & 90362.99  & 5.7  & 5.2$\times$e(-6) & 5.77$\pm$0.06   & 0.68$\pm$0.09  & 6.9$\pm$0.5   & 9.5  \\
HCCCHO          & 10$_{0}$$_{,}$$_{10}$-9$_{0}$$_{,}$$_{9}$ & 93043.31  & 24.6 & 2.4$\times$e(-5) & 5.80$\pm$0.02   & 0.58$\pm$0.05  & 3.8$\pm$0.4   & 6.1  \\
HCCCHO          & 4$_{1}$$_{,}$$_{4}$-3$_{0}$$_{,}$$_{3}$   & 99039.14  & 7.4  & 6.7$\times$e(-6) & 5.70$\pm$0.02   & 0.51$\pm$0.04  & 5.5$\pm$0.1   & 10.2  \\
\hline
H$^{13}$CCCHO   & 4$_{0}$$_{,}$$_{4}$-3$_{0}$$_{,}$$_{3}$   & 36099.33  & 4.3  & 1.3$\times$e(-6) & 5.70$\pm$0.08   & 0.82$\pm$0.17  & 0.45$\pm$0.03 & 0.52  \\
\hline
HC$^{13}$CCHO   & 4$_{0}$$_{,}$$_{4}$-3$_{0}$$_{,}$$_{3}$   & 37113.26  & 4.5  & 1.4$\times$e(-6) & 5.60$\pm$0.09   & 0.98$\pm$0.08  & 0.41$\pm$0.02 & 0.40  \\
\hline

\end{longtable}

Note:                                                               
\tablefoottext{*}{Not detected.}
\tablefoottext{a}{Transition 4$_{0}$$_{,}$$_{4}$$_{,}$$_{1}$-3$_{0}$$_{,}$$_{3}$$_{,}$$_{1}$ from o-H$_2$C$_3$O is merged with p-H$_2$C$_3$O (4$_{0}$$_{,}$$_{4}$$_{,}$$_{0}$-3$_{0}$$_{,}$$_{3}$$_{,}$$_{0}$) since they both have the same emission frequency.}
\tablefoottext{b}{Transition 5$_{0}$$_{,}$$_{5}$$_{,}$$_{1}$-4$_{0}$$_{,}$$_{4}$$_{,}$$_{1}$ from o-H$_2$C$_3$O is merged with p-H$_2$C$_3$O (5$_{0}$$_{,}$$_{5}$$_{,}$$_{0}$-4$_{0}$$_{,}$$_{4}$$_{,}$$_{0}$) since they both have the same emission frequency.}
\tablefoottext{c}{Blended with an unidentified (U) line.}
\tablefoottext{d}{Blended with p-CH$_2$CN.}
\tablefoottext{e}{Blended with the hfs of H$^{13}$CCNC.}
\tablefoottext{f}{Blended with C$_6$D.}

\begin{table*}[h!]
\centering
\caption{Rates for the DR of HCCCHSH$^+$ and C$_2$H$_3$CS$^+$.}
\begin{center}
\begin{tabular}{lllll}
\hline 
\hline
$k$ = $\alpha$$\times$($T$/300)$^{\beta}$$\times$exp(-$\gamma$/$T$) cm$^{3}$ molecule$^{-1}$  s$^{-1}$   & $\Delta$E (kJ/mol) &  $\alpha$ &  $\beta$  & $\gamma$  \\
\hline
\hline 
HCCCHSH$^+$ + e$^-$	                   $\rightarrow$ H + HCCCHS           &  -524 &  2$\times$e(-8)  & -0.7 & 0  \\ 
\textcolor{white}{HCCCHSH$^+$ + e$^-$} $\rightarrow$ H + H$_2$C$_3$S      &  -540 &  1$\times$e(-9)  & -0.7 & 0  \\ 
\textcolor{white}{HCCCHSH$^+$ + e$^-$} $\rightarrow$ H + c-$_2$C$_3$S     &  -523 &  1$\times$e(-9)  & -0.7 & 0  \\ 
\textcolor{white}{HCCCHSH$^+$ + e$^-$} $\rightarrow$ H + H + HC$_3$S      &  -183 &  1$\times$e(-7)  & -0.7 & 0  \\ 
\textcolor{white}{HCCCHSH$^+$ + e$^-$} $\rightarrow$ H + H$_2$ + C$_3$S   &  -303 &  1$\times$e(-7)  & -0.7 & 0  \\
\textcolor{white}{HCCCHSH$^+$ + e$^-$} $\rightarrow$ H + C$_2$H$_2$ + CS  &  -361 &  2$\times$e(-7)  & -0.7 & 0  \\
\textcolor{white}{HCCCHSH$^+$ + e$^-$} $\rightarrow$ SH + t-C$_3$H$_2$    &  -431 &  2$\times$e(-7)  & -0.7 & 0  \\
\hline
C$_2$H$_3$CS$^+$ + e$^-$ $\rightarrow$ H + H$_2$C$_3$S      &  -440 &  1$\times$e(-7)  & -0.7 & 0  \\ 
\textcolor{white}{C$_2$H$_3$CS$^+$ + e$^-$} $\rightarrow$ H + c-H$_2$C$_3$S     &  -423 &  1$\times$e(-8)  & -0.7 & 0  \\ 
\textcolor{white}{C$_2$H$_3$CS$^+$ + e$^-$} $\rightarrow$ H + HCCCHS           &  -424 &  1$\times$e(-8)  & -0.7 & 0  \\ 
\textcolor{white}{C$_2$H$_3$CS$^+$ + e$^-$} $\rightarrow$ H + H + HC$_3$S      &  -82  &  1$\times$e(-7)  & -0.7 & 0  \\ 
\textcolor{white}{C$_2$H$_3$CS$^+$ + e$^-$} $\rightarrow$ H + H$_2$ + C$_3$S   &  -203 &  1$\times$e(-7)  & -0.7 & 0  \\
\textcolor{white}{C$_2$H$_3$CS$^+$ + e$^-$} $\rightarrow$ H + C$_2$H$_2$ + CS  &  -260 &  3$\times$e(-7)  & -0.7 & 0  \\
\hline
\end{tabular}
\label{table:rates}
\end{center}
\end{table*}

\twocolumn

\twocolumn

\end{appendix}

\end{document}